\documentclass[reprint,prb,aps,showpacs,amsmath,amssymb]{revtex4-1}
\usepackage{graphicx}% Include figure files
\usepackage{dcolumn}% Align table columns on decimal point
\usepackage{bm}% bold math
\usepackage{float}
\begin{document}
%\preprint{Ashwin et al}
%\draft
%\draft
\title{Theoretical and Experimental Investigation on Structural, Electronic and Magnetic Properties of layered Mn$_{5}$O$_{8}$}
\author{M. R. Ashwin Kishore,$^{1}$ H. Okamoto,$^{2}$ Lokanath Patra,$^{1}$ \\
R. Vidya,$^{3}$ Anja O. Sj{\aa}stad,$^{2}$ H. Fjellv{\aa}g,$^{2}$ and P. Ravindran$^{1,2,}$}
\email{Electronic address: raviphy\texttt{@}cutn.ac.in}
\address{$^{1}$Department of Physics, Central University of Tamil Nadu, Thiruvarur, 610101, India}
\address{$^{2}$Center for Materials Science and Nanotechnology and Department of Chemistry, University of Oslo, Box 1033
Blindern, N-0315 Oslo, Norway}
\address{$^{3}$Department of Medical Physics, Anna University, Chennai, 600025, India}
\date {\today}
\begin{abstract}
{\sloppy}
We have investigated the crystal, electronic, and magnetic structure of Mn$_{5}$O$_{8}$ by means of state-of-the-art density functional theory calculations and neutron powder diffraction (NPD) measurements. This compound stabilizes in the monoclinic structure with space group C2/m where the Mn ions are in the distorted octahedral and trigonal prismatic coordination with oxygen atoms. The calculated structural parameters based on total energy calculations are found to be in excellent agreement with low temperature NPD measurements when we accounted correct magnetic structure and Coulomb correlation effect into the computation. Bond strength analysis based on crystal orbital Hamiltonian population between constituents indicating strong anisotropy in the bonding behavior which results in layered nature of its crystal structure. Using fully relativistic generalized-gradient approximation with Hubbard \textit{U} (GGA+\textit{U}) we found that the magnetic ordering in Mn$_{5}$O$_{8}$ is \textit{A-type} antiferromagnetic and the direction of easy axis is [1 0 0] in agreement with susceptibility and NPD measurements. However, the calculation without the inclusion of Hubbard\textit{U} leads to ferrimagnetic half metal as ground state contradictory to experimental findings, indicating the presence of strong Coulomb correlation effect in this material. The GGA calculations without Coulomb correction effect itself is sufficient to reproduce our experimentally observed magnetic moments in various Mn sites. We explored the electronic band characteristics using total, site-, and orbital-projected density of states and found that Mn is in two different oxidation states. A dominant Mn 3\textit{d} character observed at Fermi energy in the DOS analysis is the origin for the metallic behavior of Mn$_{5}$O$_{8}$. Our bonding analysis shows that there is a noticeable covalent bond between Mn 3\textit{d}$-$O 2\textit{p} states which stabilizes the observed low symmetric structure. We found that Mn in this material exhibits mixed valence behavior with 2+ and 4+ oxidation states reflecting different magnetic moments in the Mn sites. Our experimental findings and theoretical predications suggest that Mn$_{5}$O$_{8}$ can be classified as a strongly correlated mixed valent antiferromagnetic metal.
\end{abstract}

\maketitle
\section{Introduction}
Manganese (Mn) oxides can be considered as an interesting class of material among various transition metal oxides because they crystallize in different structures with many oxidation states (2+, 3+, and 4+ etc.)~\cite{van91,geller71,gao09,franchi07} that bring exotic magnetic behaviors. Among them, Mn$^{2+}$ has basically no preference in on coordination due to 3d$^{5}$ electronic configuration and it occupies almost indifferently crystallographic sites with coordination numbers of 4, 6, and/or 8, depending on other structural constraints. On the other hand, Mn$^{3+}$ and Mn$^{4+}$ are commonly found in octahedral sites in complex oxides. Also, Mn$^{3+}$(t$^{3}_{2g}$e$^{1}_{g}$) ion in LaMnO$_{3}$~\cite{ravi02} makes it a typical Jahn-Teller system which results in a distortion of the MnO$_{6}$ coordination octahedron along four equatorial shorter bonds and two axial longer bonds. In terms of ionic radius, the difference in the ionic radii between Mn$^{2+}$ and Mn$^{3+}$ [0.185 \AA (= 0.83 \AA - 0.645 \AA)] is larger than that between Mn$^{3+}$ and Mn$^{4+}$ [0.115 \AA (= 0.645 \AA - 0.53 \AA, based on r(O$^{2-}$) = 1.40 \AA)]~\cite{shannon76}. Namely, divalent Mn ion has somehow unique feature in oxide form, and this has important consequences on the crystal chemistry and physical properties. Binary manganese oxides such as MnO, Mn$_{3}$O$_{4}$,  Mn$_{2}$O$_{3}$, MnO$_{2}$, and Mn$_{5}$O$_{8}$ possess wide variety of technological applications due to their unique structural and physical properties. This family of compounds are quite attractive and potential candidate for catalysis, electrode materials for batteries and soft magnetic materials for transformers cores due to the presence of mixed valences of Mn atoms.~\cite{thackeray97,jeffrey99} Due to the complex interplay between orbital, spin, and lattice degrees of freedom, Mn oxides exhibits intriguing properties such as colossal magnetoresistance, metal-insulator transitions, charge as well as orbital ordering,~\cite{vidya04,vidya02} and exotic magnetic behavior.~\cite{franchi07,franchi05,imada98,mott68,harri08} Also, the magnetic properties of Mn oxides gained interest due to their unusual magnetic orderings. For example, MnO is a type-II antiferromagnetic (AFM-II) insulator below the Neel temperature of T$_{\rm N}$=118\,K,~\cite{shaked88} Mn$_{3}$O$_{4}$ show a ferrimagnetic (FiM) behavior at T$_{C}$ = 42\,K, on further cooling it exhibits spiral spin structure at 39\,K and then it transformed to canted spin array at 33\,K.~\cite{dwight60,jensen74,srini83} $\alpha$-Mn$_{2}$O$_{3}$ exhibits a complex noncollinear AFM ordering, and $\beta$-MnO$_{2}$ has a screw-type magnetic structure with ordered helical moments.~\cite{yoshi59,sato01} Among the stable binary Mn oxides, Mn$_{5}$O$_{8}$ has been reported in the literature as metastable phase.~\cite{suga91,azzoni99,rask86}

\par
\begin{figure*}
\includegraphics[scale=0.32]{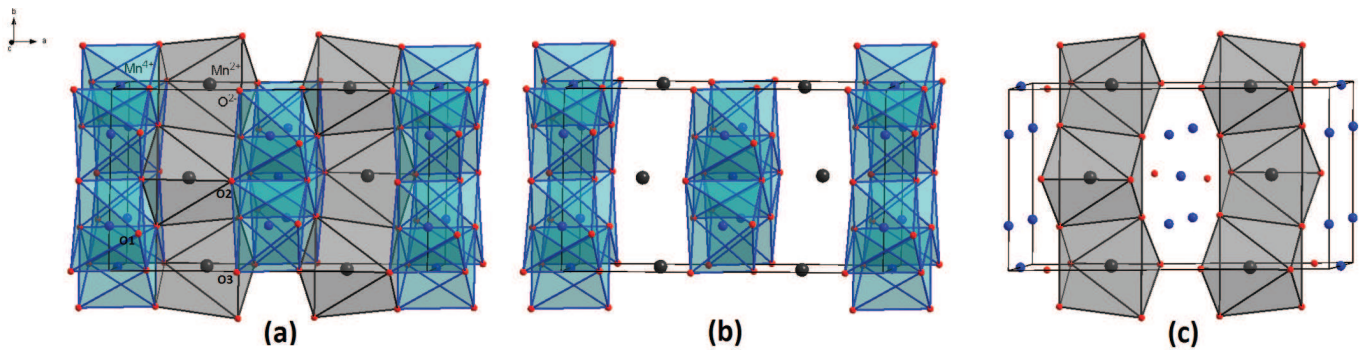}
\caption{\label{fig:struct} (Color online) (a) Polyhedral representation of Mn$_{5}$O$_{8}$ crystal structure, (b) Octahedral coordination of Mn$^{4+}$ atoms, and (c) Trigonal prismatic coordination of Mn$^{2+}$ atoms. The atom labels are given in (a).}
\end{figure*}

Mn oxides crystallize typically into tunnel$-$ and/or layered$-$ crystal structure. Single, double, or triple chains of edge-shared MnO$_{6}$ octahedra share corners with each other form tunnel structure, whereas, layer Mn oxides comprised of stack of sheets or layers of edge-shared MnO$_{6}$ octahedra, and interlayer species.~\cite{jeffrey99} Mn$_{5}$O$_{8}$ possess layered Birnessite-type structure.~\cite{auerbach04} The crystal structure reported by Oswald \emph{et al.}~\cite{oswald67} for Mn$_{5}$O$_{8}$  having compositional formula Mn$_{2}^{2+}$Mn$_{3}^{4+}$O$_{8}$ is isotypic with monoclinic Cd$_{2}$Mn$_{3}$O$_{8}$. This structure consists of distorted MnO$_{6}$ (Mn1O$_{6}^{o}$ and Mn2O$_{6}^{o}$) edge shared octahedral layers with Mn$^{4+}$ ions in the \textit{bc} plane separated by Mn$^{2+}$ ions which coordinate to six oxygen atoms forming Mn3O$_{6}$ trigonal prisms in the interlayer spaces as shown in Fig.~\ref{fig:struct}. One fourth of the cationic sites in the main octahedra sheets are vacancies which results in charge imbalance and as a consequence of that its composition becomes [Mn$_{3}^{4+}$O$_{8}$]$^{4-}$. This charge imbalance was neutralized by the Mn$^{2+}$ ions situated above and below the empty Mn$^{4+}$ sites completing the composition Mn$_{2}^{2+}$Mn$_{3}^{4+}$O$_{8}$. Since [Mn$_{3}^{4+}$O$_{8}$]$^{4-}$ layers are separated by undulating A$^{2+}$ (A = Mn, Cd, Ca) layers, one can expect low dimensional magnetic behavior in these materials.

%%%%%%%%%%%%%%%%%%% FIG 1 %%%%%%%%%%%%%
%%%%%%%%%%%%%%%%%%%%

\par
Yamamoto \textit{et al}.~\cite{yama73} reported that Mn$_{5}$O$_{8}$ orders antiferromagnetically at N\'{e}el temperature T$_{\rm N} \simeq$ 136\,K, which was highest among most of the known manganese oxides. Several other experimental studies also reported that Mn$_{5}$O$_{8}$ as an antiferromagnet with N\'{e}el temperature T$_{\rm N} \simeq$ 128\,K,~\cite{punnoose01} T$_{\rm N} \simeq$ 133\,K,~\cite{gao10} T$_{\rm N} \simeq$ 131\,K,~\cite{thota10} and T$_{\rm N} \simeq$ 126\,K.~\cite{uddin13} Noticeable difference in the reported N\'{e}el temperature is due to the finite-size effect on the antiferromagnetic transition temperature T$_{\rm N}$.~\cite{zheng05,lin07} In addition to the characteristic peak of antiferromagnetic ordering measured in Mn$_{5}$O$_{8}$ by magnetization versus temperature measurements, a sharp peak is observed at around 40\,K~\cite{punnoose01,gao10} which is believed to be associated with T$_{\rm C}$ of ferrimagnetic Mn$_{3}$O$_{4}$ as Mn$_{5}$O$_{8}$ is synthesized by the oxidation of Mn$_{3}$O$_{4}$. However, our low temperature NPD measurements were unable to detect any secondary phases and the details will be discussed below.

\par
X-ray photoemission spectroscopy (XPS) measurements on Mn$_{5}$O$_{8}$ have been made and the analysis~\cite{gao10,uddin13} confirms the two possible types of Mn with oxidation states 4+ and 2+. It may be noted that it is difficult to distinguish between Mn$^{2+}$ and Mn$^{4+}$ due to the small binding energy shift (less than ca. 1.0 eV) from Mn$^{2+}$ to Mn$^{4+}$ for the Mn 2$p$ and it will be even complicated if the Mn present in mixed valence states.~\cite{han06} Jeong \textit{et al}.~\cite{jeong15} reported that the Mn$^{3+}$ ions are involved in oxygen evolution reaction process of Mn$_{5}$O$_{8}$ nanoparticles and which is contradictory to the XPS analysis mentioned above. Therefore, apart from calculating Bond Valence Sum (BVS), we have used various theoretical tools to analyze the oxidation of manganese ions in Mn$_{5}$O$_{8}$ as we have reported earlier.~\cite{vidya2006}
\par
It is reported that the crystal structure of Mn$_{5}$O$_{8}$ is monoclinic and isostructural with Cd$_{2}$Mn$_{3}$O$_{8}$,~\cite{oswald67} Ca$_{2}$Mn$_{3}$O$_{8}$,~\cite{ansell82} Zn$_{2}$Mn$_{3}$O$_{8}$,~\cite{lecerf74} and Cu$_{2}$Mn$_{3}$O$_{8}$.~\cite{riou77} Mn$_{5}$O$_{8}$ has higher magnetic transition temperature than its isomorphous compounds such as Cd$_{2}$Mn$_{3}$O$_{8}$ (T$_{\rm N} \simeq$ 10\,K) and Ca$_{2}$Mn$_{3}$O$_{8}$ (T$_{\rm N} \simeq$ 60\,K) due to the fact that the 2+ ion is also having finite magnetic moment which results in different exchange integrals. Moreover Cd$_{2}$Mn$_{3}$O$_{8}$ and Ca$_{2}$Mn$_{3}$O$_{8}$ has lower value of exchange integral than that of Mn$_{5}$O$_{8}$ due to larger interlayer and intralayer distance between the magnetic ions which causes the higher T$_{\rm N}$ for Mn$_{5}$O$_{8}$ than those of its isomorphous compounds. The highest magnetic transition temperature in Mn$_{5}$O$_{8}$ among these isostructural compounds may be attributed to the strong interlayer exchange couplings between Mn$^{2+}$ and Mn$^{4+}$ ions with different magnetic moments.

\par
The details of magnetic ordering, the direction of easy axis, and anisotropy in the magnetic and transport properties for Mn$_{5}$O$_{8}$ are not yet identified though nanorods of Mn$_{5}$O$_{8}$ have reported recently.~\cite{gao10} To our knowledge, no theoretical studies on Mn$_{5}$O$_{8}$ have been reported in the literature. In this present study we attempt to identify the ground state magnetic structure, electronic structure, and the mixed valent behavior of manganese ions in this compound using experimental measurements and computational studies.

\section{Structural aspects and Computational Details}
\subsection{Crystal and magnetic structure}
Structurally, Mn$_{5}$O$_{8}$ crystallize in the base centered monoclinic Cd$_{2}$Mn$_{3}$O$_{8}$ $-$ type structure with space group C2/m. This structure can be considered as a layered structure with two formula units per unit cell as indicated in Fig.~\ref{fig:struct}. The input structural parameters used for the present calculations are taken from our Neutron Powder Diffraction (NPD) results obtained at 9 K (see Table.~\ref{table:npddata}).
As the convergence is very slow due to large number of atoms involved with spin polarization, spin orbit coupling, and Coulomb correlation effect included in the calculations, we have not done structural optimization for all the magnetic configurations discussed here except the ground state magnetic structure M4. Using force minimization method the atom positions are optimized with Wien2k code. Moreover, in order to find the ground state with global minima we have adopted force as well as stress minimization method implemented in VASP code for the ground state magnetic configuration and the optimized structural parameters are compared with our NPD measurements. As we have found earlier sufficiently large basis set with reasonable number of \textbf{k}-points are needed to predict reliable structural parameters for transition metal oxides,~\cite{ravi2006} we have used the energy cut off of 875 eV and the \textbf{k}-points value is 2$\times$4$\times$4 for the irreducible part of the first Brillouin zone of base centered monoclinic lattice.
%%%%%%%%%%%%%%%%%%%%% Table 1 %%%%%%%%%%%%%%%%%%%%%%%%%%%%%%%%%%
\begin{table*}
\caption{The refined unit cell dimensions, atomic coordinates, and magnetic moments at various Mn sites for Mn$_{5}$O$_{8}$, derived from Rietveld refinement of NPD data 9\,K well below the magnetic transition temperature are listed in this table. The crystal structure is found to be base centered monoclinic with space group C2/m. The calculated standard deviations are given in parentheses.\\
\emph{a} = 10.325(2) \AA, \emph{b} = 5.7181(7) \AA, \emph{c} = 4.8594(6) \AA, $\beta$ = 109.63(2)$^{o}$\\
R$_{wp}$ = 4.86\%, R$_{P}$ = 3.67\%, $\chi^{2}$ = 2.08 }
\begin{ruledtabular}
\begin{tabular}{lcccccccccc}
Atom & & & & & & & Moment** & & & \\
\hline\\
& Wyckoff position & $x$ & $y$ & $z$ & Occupancy & B$_{iso}$(\AA$^{2})^{*}$ & M$_{x}(\mu_{B})$ & M$_{y}(\mu_{B})$ & M$_{z}(\mu_{B})$ & M$_{tot}(\mu_{B})$ \\
\hline
Mn1	& 2c (0 0 1/2)	& 0	& 0	& 1/2	& 1	& 0.2(2) & 2.3(2) & 0 & 1.8(2) & 2.4(2) \\
Mn2	& 4g (0 y 0)	& 0	& 0.247(4)& 0 &	1 &	0.2(2) & 2.3(2) &	0 &	1.8(2) & 2.4(2) \\
Mn3	& 4i (x 0 z)	& 0.729(2) & 0 & 0.665(5) &	1 &	0.2(2) & 4.02(9) &	0 &	1.7(3) & 3.8(2) \\
O1	& 8j (x y z)	& 0.114(2)	& 0.222(2) & 0.395(2) &	1 &	1.7(2) \\
O2	& 4i (x 0 z)	& 0.114(2)	& 0	& 0.922(5) & 1 & 1.7(2) \\
O3	& 4i (x 0 z)	& 0.591(2)	& 0	& 0.872(4)	& 1	& 1.7(2)
\label{table:npddata}
\let\thefootnote\relax\footnotetext{*Isotropic atomic displacement parameters were constrained at same values for respectively, Mn and O.\\
**Magnetic moments were constrained at same values for Mn1 and Mn2.}
%\footnote{*Isotropic atomic displacement parameters were constrained at same values for respectively, Mn and O.\\
%**Magnetic moments were constrained at same values for Mn1 and Mn2.}
\end{tabular}
\end{ruledtabular}
\end{table*}
%%%%%%%%%%%%%%%%%%%%% End Table 1 %%%%%%%%%%%%%%%%%%%%%%%%%%%%%%%%%%

\par
The overall crystal structure of Mn$_{5}$O$_{8}$ can be described as infinite elemental sheets of [Mn$_{3}^{4+}$O$_{8}$]$^{4-}$ where the Mn$^{4+}$ ions are held together by Mn$^{2+}$ ions. This material is a mixed valence system with two different types of Mn$^{4+}$ ions and one type of Mn$^{2+}$ ion. Also, the oxygen ions are located into three different sites in Mn$_{5}$O$_{8}$. One of these Mn$^{4+}$ (Mn1$^{o}$) ion is in 2\emph{c} site (C$_{2h}$ symmetry) coordinated octahedrally to four planar O atoms at 1.921 \AA\ and two apical O atoms at 1.937 \AA\ where the average Mn$-$O distance is 1.926 \AA. Angles between Mn and O within the plane in the Mn1O$_{6}^{o}$ octahedra range from 85${^o}$ to 95${^o}$ and that between Mn and apical oxygen is 180${^o}$. This has lot of influence on the magnetic exchange interaction in this system. Two Mn$^{4+}$ (Mn2$^{o}$) atoms are in 4g sites (C$_{2}$ symmetry) coordinated octahedrally with highly distorted octahedra where two O atoms at 1.867 \AA\ and two at 1.913 \AA\ and two at 1.971 \AA\ with an average Mn$-$O distance of 1.917 \AA. The O$-$Mn$-$O angles within this Mn2O$_{6}^{o}$ distorted octahedra varies between 81.9${^o}$ to 96.4${^o}$ when the O atoms are adjacent to each other and varies between 168.7${^o}$ to 177.6${^o}$ when the O atoms are non-adjacent to each other. Mn$^{2+}$ ions (Mn3$^{t}$) form a trigonal-prismatic coordination with six oxygen atoms in which three are from one octahedral layer and three are from the next octahedral layer. As described in Fig.~\ref{fig:struct}, O(1) is coordinated by two Mn$^{4+}$ and two Mn$^{2+}$ ions, O(2) is coordinated by three Mn$^{4+}$ and one Mn$^{2+}$ ions, and O(3) is coordinated by two Mn$^{4+}$ and one Mn$^{2+}$ ion. Due to this difference in the coordination of O atom with the Mn ions bring different magnetic moment in various manganese sites.

\par
Based on our experimental results we have proposed four different magnetic structure models for Mn$_{5}$O$_{8}$ namely M1, M2, M3, and M4 (see Fig.~\ref{fig:magstruct}). So, in order to identify the ground state among these proposed four magnetic structures we have done \emph{ab initio} total energy calculation for all these four models including spin-orbit coupling as well as Coulomb correlation effects. In M1 magnetic configuration, it may be noted that the ferromagnetic (FM) coupling is present within the layers and antiferromagnetic (AFM) coupling is present between the layers within \textit{ac} plane where spin is oriented along [1 0 0] direction. In M2 magnetic configuration, again there is a FM coupling within the layers and an AFM coupling between the layers with spin orientation is along [0 1 0], In the case of M3 magnetic configuration, there is a FM coupling within the layers and an AFM interaction between the layers along \textit{b} axis where spin orientation is along [0 1 0]. For the M4 magnetic configuration, FM interaction is present within the layers and an AFM interaction is present between the layers with spin orientation is along [1 0 0] as shown in Fig.~\ref{fig:magstruct}. In both M1 and M2 magnetic configurations, the magnetic moment at the Mn1 and Mn2 sites are smaller than that in the Mn3 sites. The lower moments in the Mn1 and Mn2 are antiferromagnetically coupled with the higher moment in the Mn3 site bringing ferrimagnetic ordering with finite moment. However, the magnetic moments in the models M3 and M4 are exactly canceling due to perfect antiferromagnetic ordering.

\subsection{Computational details for the full-potential linear augmented plane-wave (FP-LAPW) calculations}
Calculations were performed using standard full-potential linear augmented plane-wave method based on density functional theory (DFT) as implemented in the Wien2k code.~\cite{blaha01} We have used the generalized gradient approximation (GGA) of Perdew-Burke-Ernzerhof for the exchange-correlation functional.~\cite{perdew96} The muffin tin sphere radii (R$_{MT}$) values for Mn and O atom were taken to be 1.87 and 1.61 a.u, respectively. The Plane wave cut-off parameters were decided by R$^{min}_{MT}$K$_{max}$ = 7 (the product of the smallest of the atomic sphere radii R$_{MT}$ and the plane wave cut-off parameter K$_{max}$) and 400 \textbf{k}-points were used over the irreducible part of the first Brillouin zone (IBZ). We have also included both Mott-Hubbard parameter \textit{U}, and spin-orbit coupling (\textit{SO}) in the calculations to account for correlation effect (GGA+\textit{U}) and relativistic effect(GGA+\textit{SO}) respectively. The value of \textit{U}$_{\rm eff}$=\textit{U}-\textit{J} (\textit{U} and \textit{J} are on-site Coulomb and exchange interaction, respectively) chosen for this calculation is 5\,eV which is normally used for manganese oxides to account for Coulomb correlation effects.

\subsection{Computational details for Vienna \emph{ab-initio} simulation package (VASP) calculations}
Structural optimizations were done using VASP code~\cite{kresse96} within the projector augmented wave (PAW) method. The generalized gradient approximation proposed by Perdew, Burke and Ernzrhof (GGA-PBE)~\cite{perdew96} was used for the exchange and correlation functional. IBZ was sampled with a 2$\times$4$\times$4 Monkhorst-Pack \textbf{k}-point mesh centered at $\Gamma$ point in the Brillouin zone. As mentioned above, the optimized structural parameters for transition metal oxides are very sensitive to the size of the basis set. We have used very large energy cut-off of 875 eV for the present calculations. The force minimization steps were continued until the maximum Hellmann-Feynman forces acting on each atom is less than 0.01\,eV/{\AA}. Also, the pressure in the unit cell was kept below 1 kbar.

\section{Experimental details}
Polycrystalline sample of Mn$_{5}$O$_{8}$ was synthesized by previously reported wet chemical reaction route.~\cite{gao2009} Prior to NPD measurements, the synthesized sample was characterized by X-ray diffraction and found that there is no detectable impurities. Neutron powder diffraction (NPD) data at 298 K, 70 K, and 9 K were collected with the PUS two-axis diffractometer at the JEEP II reactor, Kjeller, Norway\cite{hauback00}. The powder sample was kept in cylindrical sample holders. Monochromatized neutrons of wavelength 1.5556 \AA\, were obtained by reflection from Ge (311). NPD data was measured in the 10.00$^{o}$ $\leq$ 2$\theta$ $\leq$ 129.95$^{o}$ with the scan-step of 0.05$^{o}$. The simulation and Rietveld methods\cite{rietveld69} were performed with Fullprof code.\cite{rodriguez93} NPD data and reflections from the Displex cryostat at 46.00$^{o}$ $\leq$ 2$\theta$ $\leq$ 46.90$^{o}$, 76.90$^{o}$ $\leq$ 2$\theta$ $\leq$ 77.60$^{o}$, and 115.90$^{o}$ $\leq$ 2$\theta$ $\leq$ 117.20$^{o}$ in all collected NPD data were excluded prior to the refinements. Scale factor, zero point, pseudo-Voigt profile parameters, unit-cell dimensions, positional parameters, together with four isotropic displacement factors, and the magnetic moments of manganese, were entered into the final least-squares refinement.

%%%%%%%%%%%%%%%%%%%  FIG 1 %%%%%%%%%%%%%
\begin{figure}[h]
\includegraphics[scale=0.32]{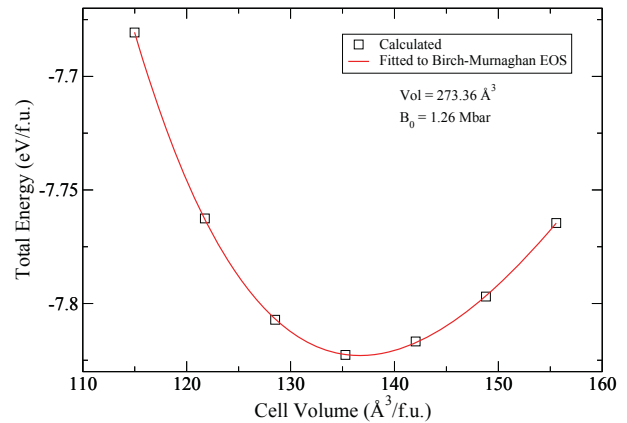}
\caption{\label{fig:eos} Calculated total energy versus cell volume for Mn$_{5}$O$_{8}$ in the ground state M4 configuration.}
\end{figure}
%%%%%%%%%%%%%%%%%%%%

%%%%%%%%%%%%%%%%%%%  FIG 2 %%%%%%%%%%%%%
\begin{figure}[h]
\includegraphics[scale=0.15]{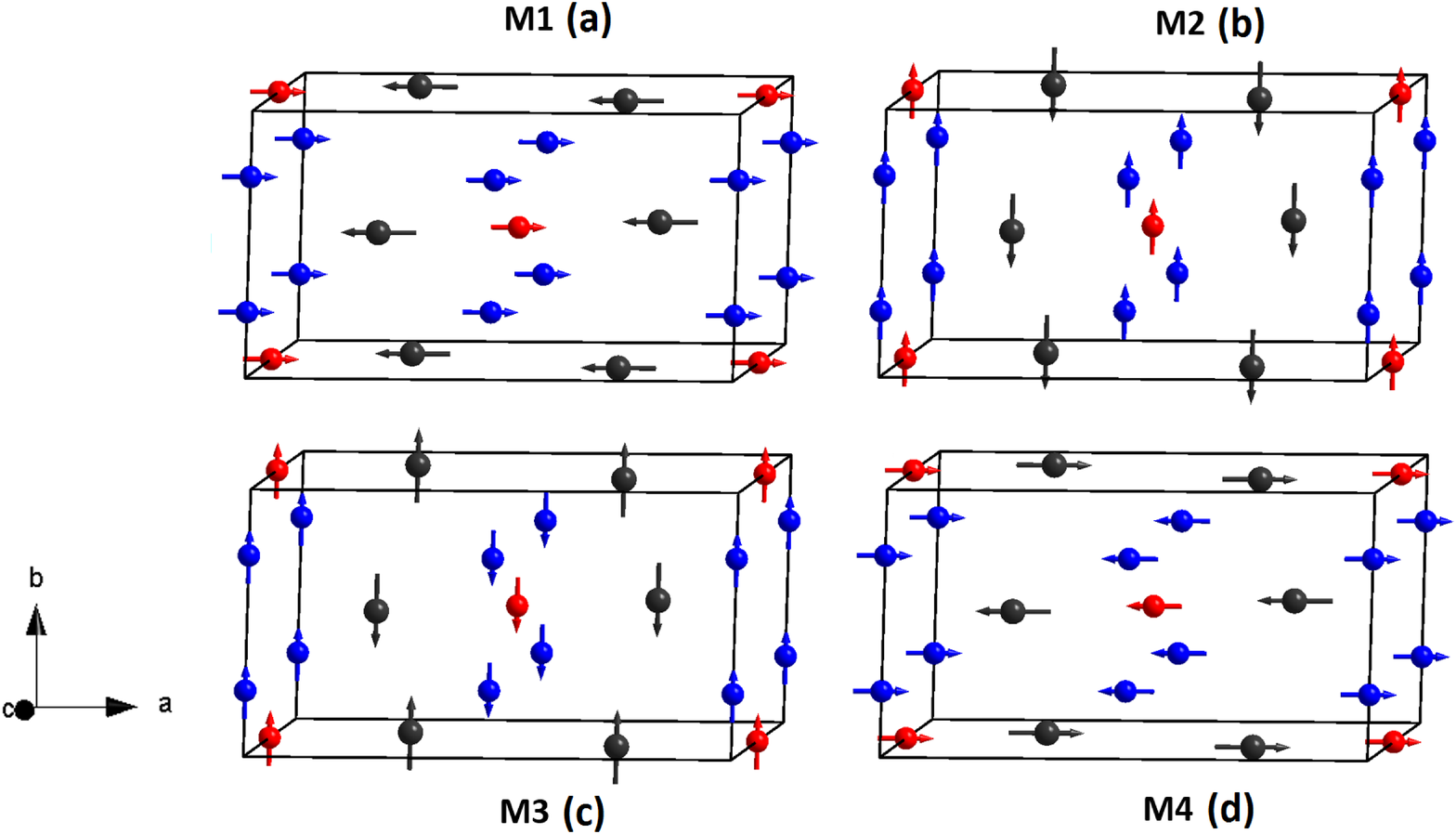}
\hspace{0.10in}
\includegraphics[scale=0.45]{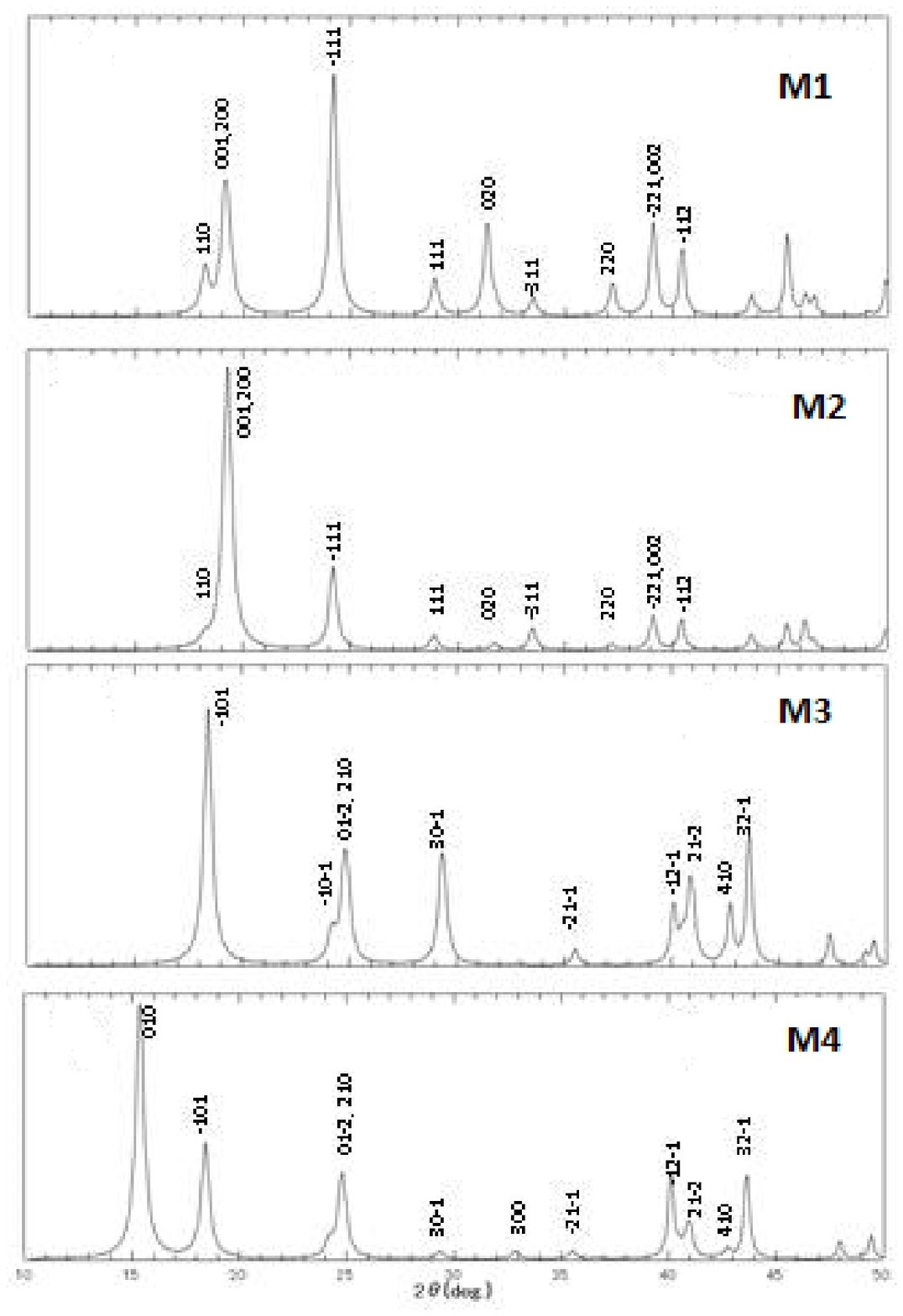}
\caption{\label{fig:magstruct} (Color online) The proposed magnetic structures M1, M2, M3, and M4 based on experimental measurements are shown as (a),(b),(c), and (d), respectively. The manganese ions with different spin moment are denoted as Mn1(Red), Mn2(Blue), and Mn3(Grey) in these figures. The corresponding simulated Neutron Powder Diffraction patterns for Mn$_{5}$O$_{8}$ in these four magnetic models are given in the lower panel.}
\end{figure}
%%%%%%%%%%%%%%%%%%%%

\section{Results and Discussion}
\begin{figure}[h]
\includegraphics[scale=0.38]{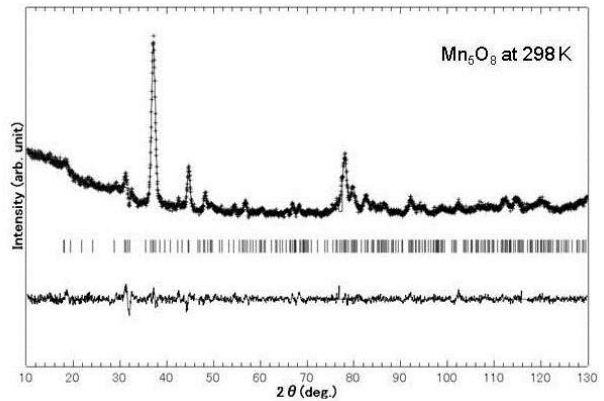}
\caption{\label{fig:npd298} Experimental, calculated, and difference NPD patterns of Mn$_{5}$O$_{8}$ at 298 K. Bars denote Bragg positions; space group C2/m, $\lambda$ = 1.5556 \AA.}
\end{figure}

\begin{figure}[h]
\includegraphics[scale=0.27]{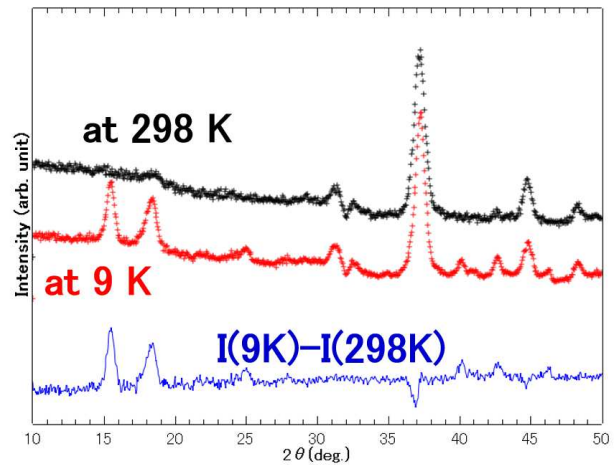}
\caption{\label{fig:npddiff} (Color online) Experimental NPD data at 298 K and 9K, and corresponding differential curve.}
\end{figure}

Before discussing the magnetic structure of Mn$_{5}$O$_{8}$, we would like to describe NPD data at 298\,K based on the result of Rietveld analysis as shown in Fig.~\ref{fig:npd298}. Careful indexing of all observed peaks in the collected data confirms the previously proposed systematic extinction scheme,~\cite{oswald65} corresponding to space group of C2/m (No. 12) ($a$ = 10.372 \AA, $b$ = 5.733 \AA, $c$ = 4.860 \AA, and $\beta$ = 109.4$^{o}$).
Present NPD pattern at 9\,K show extra intensity compared to that at 298\,K, which can be expected from magnetic reflection arising from magnetic ordering which will take place around T$_{N}$ = 136 K as shown in Fig.~\ref{fig:npddiff}. To enhance the magnetic reflections, in Figure~\ref{fig:npddiff} we have also plotted the differential intensity between 298\,K and 9 K [I(9\,K) - I(298\,K)]. We have assumed four possible AFM models in the crystal structural dimension 1 $\times$ 1 $\times$ 1, which are shown in Figs.~\ref{fig:magstruct} together with corresponding simulated scattering patterns. It may be noted that we have used the ratio of the magnetic moments for octahedral Mn$^{4+}$ and trigonal prismatic Mn$^{2+}$ of 2:1 for the present NPD pattern simulation. For these simulation we have also used the peak profile function yielded from Rietveld analysis at 298\,K.

\par
\begin{figure}[h]
\includegraphics[scale=0.47]{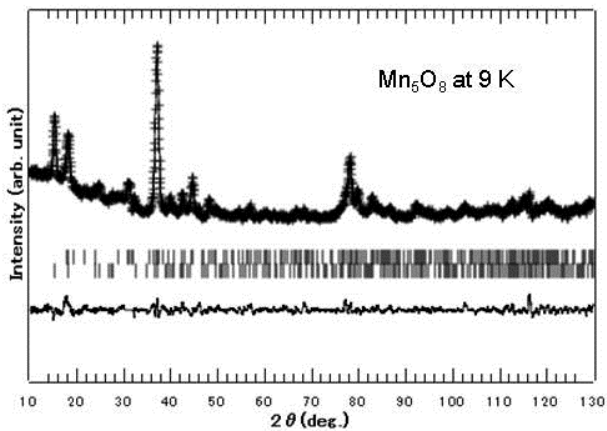}
\caption{\label{fig:npd} Experimental, calculated, and difference NPD patterns of Mn$_{5}$O$_{8}$ at 9 K. Upper and lower bars denote Bragg positions of nuclei and magnetic reflections respectively. space group C2/m, $\lambda$ = 1.5556 \AA.}
\end{figure}

As seen in Fig.~\ref{fig:npddiff}, there is a magnetic contribution outside of the Bragg positions, e.g. (0 1 0) and (-1 0 1) arising from the chemical unit cell with space group C2/m in the NPD pattern taken at 9\,K. It is worth noting that there is no magnetic super reflections at (hkl) with non-integer value of h, k, and/or l, and hence one can rule out the possibility of forming complicated modulated magnetic ordering in Mn$_{5}$O$_{8}$. We have found that the magnetic model associated with M4 alone is compatible with the measured low temperature diffraction data with observed magnetic reflections. In this model, the magnetic moments at the Mn1, Mn2, and Mn3 (and symmetry related positions) sites exist only in the \textit{ac} plane, and hence no magnetic component was found along the \textit{b}-axis. The components of magnetic moment along \textit{a}- and \textit{c}-axes, i.e. m$_{x}$ and m$_{z}$, were refined using Rietveld refinement for Fig.~\ref{fig:npd} and that for Mn1, Mn2, and Mn3 sites are found to be 2.4 $\mu_{B}$, 2.4 $\mu_{B}$, and  3.8 $\mu_{B}$, respectively.  The refined magnetic moment data are tabulated in Table~\ref{table:npddata} for the 9\,K NPD pattern.
In order to identify the equilibrium structural parameters we have performed the total energy calculation for Mn$_{5}$O$_{8}$ as a function of volume for the fully relaxed M4 structure as shown in Fig.~\ref{fig:eos}. The calculated equilibrium
structural parameters for Mn$_{5}$O$_{8}$ using VASP and Wien2k codes are compared with our low temperature NPD data in Table~\ref{table:lattpara}. The equilibrium volume obtained from Fig.~\ref{fig:eos} using Birch-Murnaghan equation of states (EOS) is overestimated only 1.1\% compared with the experimental values and hence the calculated lattice parameters are in good agreement with our corresponding experimental values. From the EOS fit we have calculated the bulk modulus and its pressure derivative for Mn$_{5}$O$_{8}$ and are given in Table~\ref{table:lattpara}. Haines \emph{et al.}~\cite{haines95} reported that bulk modulus and its pressure derivative of $\beta$-MnO$_{2}$ is 3.28 Mbar and 4 respectively. Jeanloz \emph{et al.}~\cite{jeanloz87} measured experimentally, the bulk modulus and its pressure derivative of MnO is 1.70 MBar and 4.8, respectively. Using a linear muffin-tin orbital atomic sphere approximation (LMTO-ASA) approach, Cohen \emph{et al.}~\cite{cohen97} calculated the bulk modulus and its pressure derivative of MnO is 1.96 Mbar and 3.9 respectively. Our calculated bulk modulus for Mn$_{5}$O$_{8}$ is smaller than MnO and $\beta$-MnO$_{2}$ hence experimental high pressure studies are needed to confirm our predictions.

\par
\begin{figure*}
\begin{minipage}{\textwidth}
\vspace{0.10in}
\includegraphics[scale=0.45]{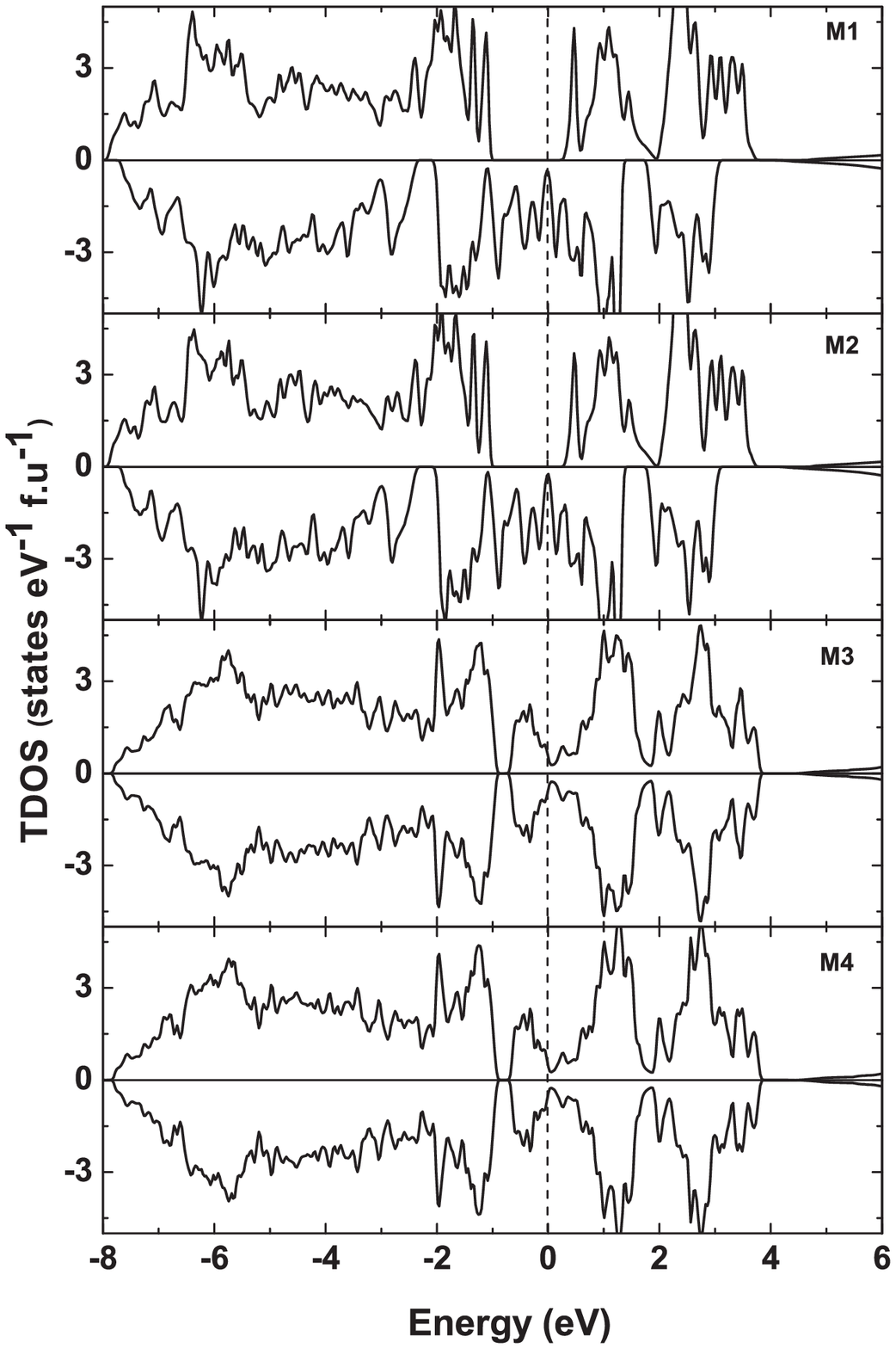}
\vspace{0.10in}
\includegraphics[scale=0.45]{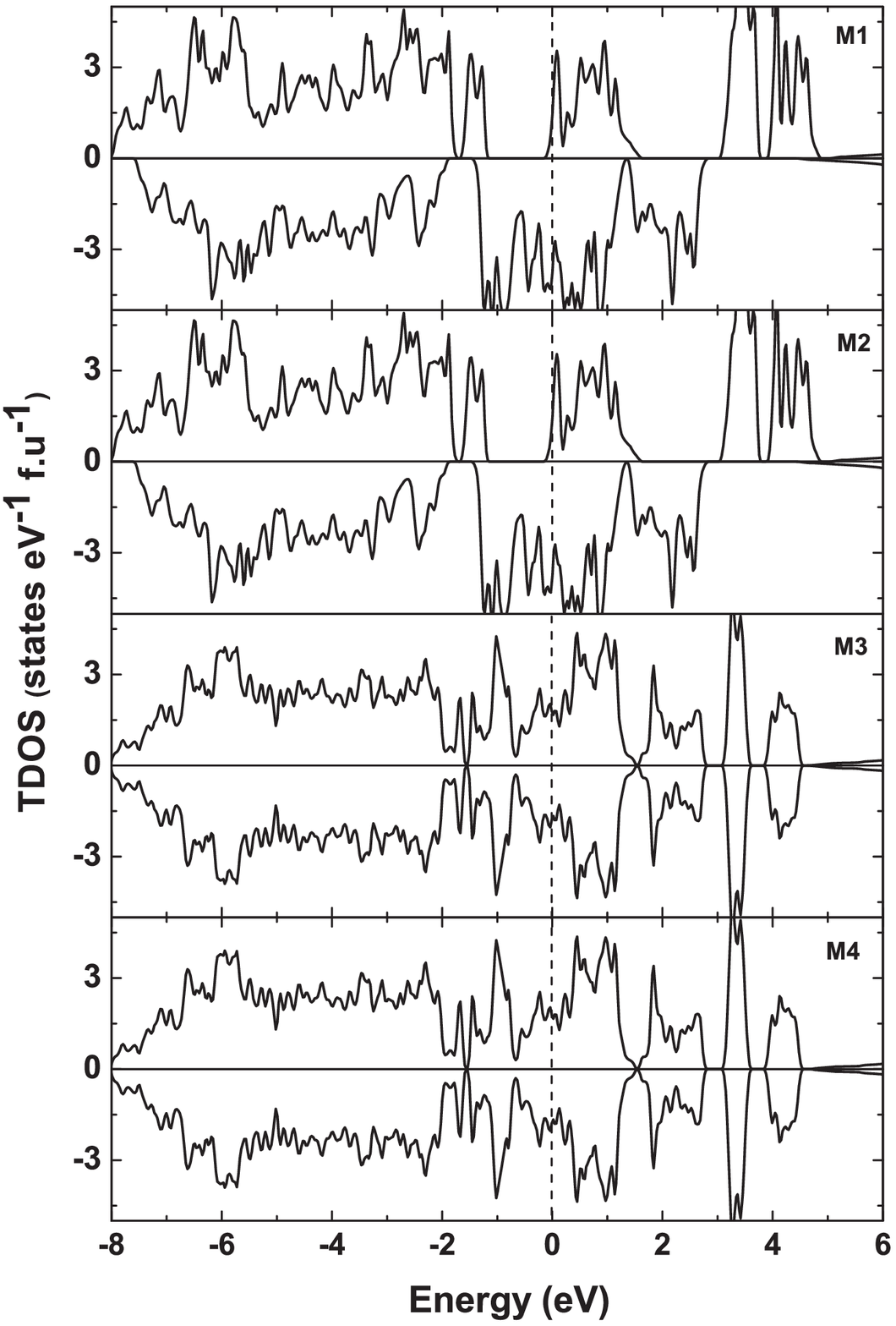}
\caption{\label{fig:totdos} The calculated total DOS for Mn$_{5}$O$_{8}$ in four different magnetic configuration M1, M2, M3, and M4 described in the text. The DOS calculated using GGA+\textit{SO} and GGA+\textit{SO}+\textit{U} are given in left and right panel, respectively. The Fermi level is set to zero.}
\end{minipage}
\end{figure*}

The calculated total density of states (DOS) for the proposed magnetic structures M1, M2, M3, and M4 are shown in Fig.~\ref{fig:totdos}. It may be noted that our GGA+\textit{SO} calculations predict M1 magnetic configuration as ground state with ferrimagnetic half metallic behavior. A finite DOS is present in the minority-spin channel at E$_{F}$ indicating metallic behavior. However, in the majority spin channel a band gap of 1.283 eV opens up resulting half metallicity in the system. It may be noted that our GGA+\textit{SO} calculations predict M1 magnetic configuration as ground state with ferrimagnetic half metallic behavior. The calculated magnetic moments are listed in Table~\ref{table:moment} which is in good agreement with the experimental neutron diffraction study. We found that the total DOS distribution and the spin moment are almost the same for M1 and M2; and also M3 and M4 and this is consistent with our experimental observation. As the magnetic ground state is found to be ferrimagnetic, whereas the experimentally observed magnetic ground state is antiferromagnetic, we concluded that GGA+\textit{SO} calculations unable to predict correct magnetic ground state in Mn$_{5}$O$_{8}$. It is well known that the transition metal oxides are usually have strong Coulomb correlation which is not accounted in our GGA+\textit{SO} calculations. Hence, we have made total energy calculation by accounting Coulomb correlation effects through GGA+\textit{SO}+\textit{U}. Interestingly when we include Coulomb correlation effect in our calculation the total DOS obtained for M1 configuration having narrow \textit{d} band states is moved towards lower energy bringing metallic states instead of half metallic behavior. Further, in major spin channel we have found that sharp peak in the DOS curve indicating instability in the system. As a result, the M1 configuration is energetically unfavorable compared with the M4 configuration. In the case of M4 configuration the magnetic moments are exactly canceled between various Mn sublattices bringing perfect antiferromagnetic ordering with metallic behavior which is consistent with experimental observations. So, we can classify that Mn$_{5}$O$_{8}$ as a strongly correlated antiferromagnetic metal. As the orbital moments obtained from relativistic spin-polarized calculations are usually smaller than corresponding experimental value,~\cite{ravindran01} we have calculated the orbital moments (see Table~\ref{table:orbmoment}) using orbital polarization correction proposed by Brooks and Erisson \emph{et al.}~\cite{brooks85,eriksson89} We found that the calculated orbital moments obtained from \textit{SO} calculation and \textit{SO}+OP calculations are not changed much 0.0072 $\mu_{B}$ and also the estimated orbital moments are small 0.030 $\mu_{B}$ as usually expected in transition metal compounds where spin-orbital coupling is expected to be weak.

%%%%%%%%%%%%%%%%%%%%%%%%% Table 2 %%%%%%%%%%%%%%%%%%%%%%%%%%%%%%%%%%
\begin{table}[b]
\caption{Calculated orbital magnetic moment ($\mu_{B}$) at various Mn sites for Mn$_{5}$O$_{8}$ without (\emph{SO}) and with orbital-polarization
corrections (\emph{SO}+OP).}
%\scriptsize
\begin{ruledtabular}
\begin{tabular}{lccc}
&  Orbital moment (Mn1/Mn2/Mn3) \\
\hline
M4 (\emph{SO})& -0.018/-0.023/0.006  \\
M4 (\emph{SO}+OP)& -0.022/-0.030/0.007
\label{table:orbmoment}
\end{tabular}
\end{ruledtabular}
\end{table}
%%%%%%%%%%%%%%%%%%%%% End Table 2 %%%%%%%%%%%%%%%%%%%%%%%%%%%%%%%%%%

%%%%%%%%%%%%%%%%%%%%% Begin Table 3 %%%%%%%%%%%%%%%%%%%%%%%%%%%%%%%%%%
\begin{table*}
\caption{Optimized structural parameters (\emph{a,b,c} in {\AA}, $\beta$ in deg, equilibrium volume V$_{0}$ in {\AA}$^{3}$), and bulk modulus (B$_{0}$ in Mbar) and its pressure derivative (B$_{0}$$^{'}$) for Mn$_{5}$O$_{8}$ in ground state M4 configuration from VASP and Wien2k$^{a}$ calculations. Values given in parenthesis refer to our low temperature NPD measurements. The space group is C2/m and Z=2.}
\begin{ruledtabular}
\begin{tabular}{lccccccc}
Unit cell & Atom & Site & $x$ & $y$ & $z$ & B$_{0}$ & B$_{0}$$^{'}$\\
\hline
\emph{a}=10.3228(10.3250)    & Mn1  & 2c & 0.0000(0.0000) & 0.0000(0.0000) & 0.5000(0.5000) &	1.26 & 5.10\\
\emph{b}=5.7436(5.7181)& & & 0.0000 & 0.0000 & 0.5000$^{a}$  & &\\
\emph{c}=4.8887(4.8594)      & Mn2  & 4g & 0.0000(0.0000) & 0.2600(0.2470) & 0.0000(0.0000) & & \\
$\beta$=109.43(109.63) & & & 0.0000 & 0.2603 & 0.0000$^{a}$ & &\\
V$_{0}$=273.36(270.24)     & Mn3  & 4i & 0.7163(0.7290) & 0.0000(0.0000) & 0.6576(0.6550) & & \\
& & & 0.7173 & 0.0000 & 0.6595$^{a}$ & &\\
& O1   & 8j & 0.1137(0.1140) & 0.2272(0.2220) & 0.3984(0.3950) & &\\
& & & 0.1134 & 0.2274 & 0.3995$^{a}$ & &\\
      & O2   & 4i & 0.1054(0.1140) & 0.0000(0.0000) & 0.9118(0.9220) & &\\
& & & 0.1049 & 0.0000 & 0.9119$^{a}$ & &\\
                      & O3   & 4i & 0.6064(0.5910) & 0.0000(0.0000) & 0.9247(0.8720) & &\\
& & & 0.6062 & 0.0000 & 0.9252$^{a}$ & &\\
\label{table:lattpara}
\end{tabular}
\end{ruledtabular}
\end{table*}
%%%%%%%%%%%%%%%%%%%%% End Table 3 %%%%%%%%%%%%%%%%%%%%%%%%%%%%%%%%%%

%%%%%%%%%%%%%%%%%%%  FIG 2 %%%%%%%%%%%%%
%%%%%%%%%%%%%%%%%%%%

%%%%%%%%%%%%%%%%%%%  FIG 2 %%%%%%%%%%%%%
%%%%%%%%%%%%%%%%%%%%

%%%%%%%%%%%%%%%%%%%  FIG 2 %%%%%%%%%%%%%
%%%%%%%%%%%%%%%%%%%%

%%%%%%%%%%%%%%%%%%%  FIG 2 %%%%%%%%%%%%%
%\begin{figure}[h]
%\includegraphics[scale=0.47]{fig/npd70.eps}
%\caption{\label{fig:npd70} Experimental, calculated, and difference NPD patterns of Mn$_{5}$O$_{8}$ at 70 K. Upper and lower bars denote Bragg positions of, respectively, nuclei and magnetic reflections; space group C2/m, $\lambda$ = 1.5556 \AA.}
%\end{figure}
%%%%%%%%%%%%%%%%%%%%

%%%%%%%%%%%%%%%%%%%%%%%%% Table 1 %%%%%%%%%%%%%%%%%%%%%%%%%%%%%%%%%%
\begin{table}[b]
\caption{The total magnetic moment, site projected magnetic moment at various Mn sites,
 and the relative total energy $(\Delta E)$ of various magnetic configurations with respect to the ground state are listed in this table.}
%\scriptsize
\begin{ruledtabular}
\begin{tabular}{lcccc}
& Magnetic & Total mom. & Mn mom. & Total \\
& model & ($\mu_{B}$/f.u) & Mn1/Mn2/Mn3 & energy\\
& & & ($\mu_{B}$/atom) & (meV/atom) \\
\hline
& M1 & -0.99 & 2.26/2.30/-3.84 &  0 \\
GGA & M2 & -0.99  &2.26/2.30/-3.84  & 78.21  \\
+\textit{SO} & M3 & 0 & 2.72/2.41/-3.83 & 52.32  \\
& M4 & 0 & 2.72/2.41/-3.83 &  52.29 \\
\hline
& M1 & -1 & 2.51/2.59/-4.21 & 84.29  \\
GGA & M2 & -1 & 2.51/2.59/-4.21 & 84.44 \\
+\textit{SO}+\textit{U} & M3 & 0 & 3.05/2.69/-4.18 & 0.04 \\
& M4 & 0 & 3.05/2.69/-4.18 & 0 \\
\hline
Exp.&  & 0 & 2.4/2.4/-3.8 \\
\label{table:moment}
\end{tabular}
\end{ruledtabular}
\end{table}
%%%%%%%%%%%%%%%%%%%%% End Table 1 %%%%%%%%%%%%%%%%%%%%%%%%%%%%%%%%%%

\par
\begin{figure}[h]
%\begin{minipage}{\textwidth}
%\hspace{-0.40in}
%\includegraphics[width=.45\textwidth]{fig/fig5.eps}
%\vspace{0.10in}
\includegraphics[width=.48\textwidth]{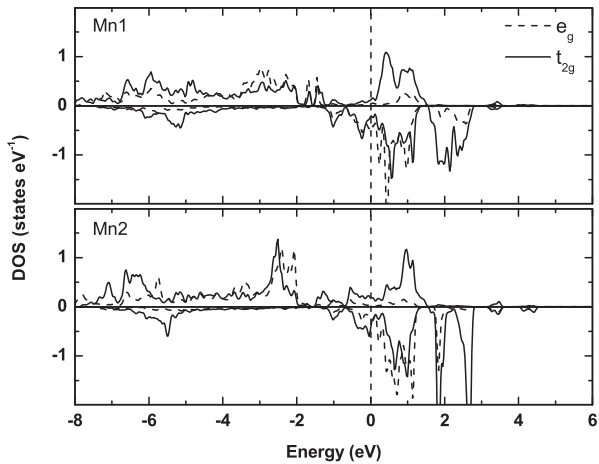}
\caption{\label{fig:orbdos} The calculated orbital decomposed DOS for Mn1 and Mn2 sites of Mn$_{5}$O$_{8}$ in the ground state antiferromagnetic configuration
obtained from GGA+\textit{SO}+\textit{U}. The Fermi level is set to zero.}
%\end{minipage}
\end{figure}

The orbital-projected DOS for Mn 3\textit{d} electrons in the ground state M4 magnetic configuration within GGA+\textit{SO}+\textit{U} is shown in Fig.~\ref{fig:orbdos}. It is well known that the Mn \textit{d} orbitals in octahedral coordination split into t$_{2g}$ triplet (d$_{xy}$, d$_{xz}$, d$_{yz}$) and e$_{g}$ doublet (d$_{x^{2}-y^{2}}$, d$_{z^{2}}$) by the cubic crystal field. As both Mn1 and Mn2 are in octahedral coordination with oxygen their magnetic moments in these two sites can be analyzed easily using orbital projected DOS. For Mn$^{4+}$ oxidation state, there will be totally three \textit{d} electrons those will occupy the majority spin t$_{2g}$ states and the e$_{g}$ states will be empty in the high spin configuration. So one can expect 3 $\mu_{B}$ per Mn sites in a pure ionic picture. In conformity with above view our calculated DOS shows that the e$_{g}$ states is almost empty in the valence band and also in the minority spin channel the t$_{2g}$ states is negligibly small confirming the high-spin (HS) state of Mn$^{4+}$ ions. However the calculated magnetic moments in the Mn2 site is lower than 3 $\mu_{B}$ (2.69$\mu_{B}$) indicating that there is a substantial covalent bond between Mn2$-$O. It may be noted that the electrons in solids may participate either bonding or magnetism. Consistent with the above point due to this covalency effect the average bond length between Mn2$-$O(1.917\,\AA) is smaller than that between Mn1$-$O(1.926\,\AA). Due to short Mn2$-$O distance there is substantial induced moment of 0.027 $\mu_{B}$ per atom present in the oxygen sites around Mn2. However magnetic moment in oxygen sites around Mn1 is small 0.006 $\mu_{B}$ per atom. The calculated exchange splitting energy for Mn1 and Mn2 are 2.01 eV and 1.9 eV, respectively correlating linearly with corresponding magnetic moment. So, the larger exchange splitting at the Mn1 site could explain why larger moments in this site compared with Mn2 site. Due to the Coulomb correlation effect \textit{d} states get localized and hence magnetic moment is increased in GGA+\textit{SO}+\textit{U} calculation as indicated in Table~\ref{table:moment}. As the pseudocubic crystal field operating in the MnO$_{6}$ octahedra, the electronic states at the Fermi level displays both Mn(e$_{g}$) and Mn(t$_{2g}$) bands. The calculated energy difference between M3 and M4 models is less than 30 $\mu$ eV only in the GGA+\textit{SO} calculation and that is increased slightly to 40 $\mu$ eV when we account correlation effect into the calculation. However, the GGA+\textit{SO}+\textit{U} calculation correctly predicted experimentally observed antiferromagnetic ground state with magnetic configuration of M4 suggesting the presence of strong Coulomb correlation effect in this material.

%%%%%%%%%%%%%%%%%%%  FIG 3 %%%%%%%%%%%%%
%%%%%%%%%%%%%%%%%%%%

%%%%%%%%%%%%%%%%%%%  FIG 4 %%%%%%%%%%%%%
%%%%%%%%%%%%%%%%%%%%

%%%%%%%%%%%%%%%%%%%  FIG 6 %%%%%%%%%%%%%
\begin{figure}[h]
%\begin{minipage}{\textwidth}
%\hspace{0.10in}
\includegraphics[width=.46\textwidth]{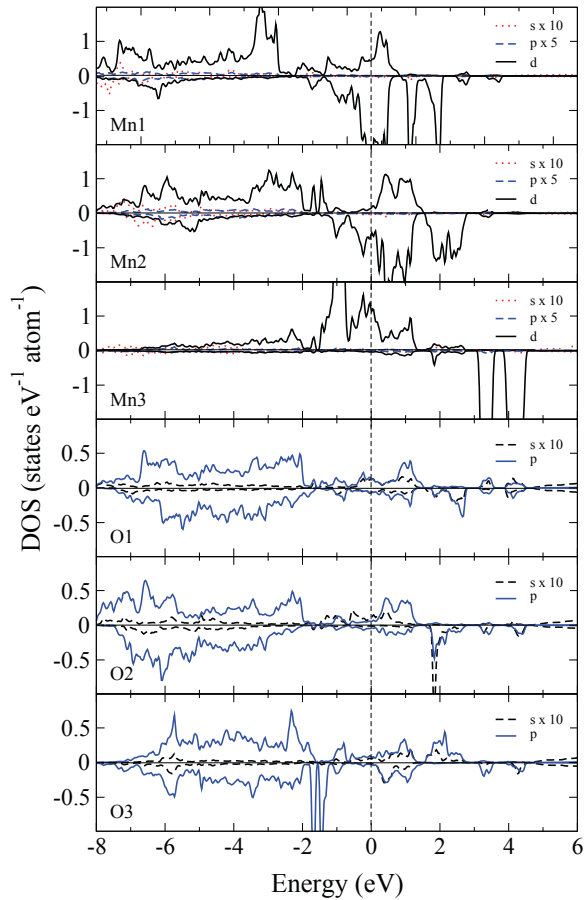}
\caption{\label{fig:lprojdos} (Color online) Calculated orbital-projected DOS for Mn and O sites of Mn$_{5}$O$_{8}$ in the ground-state M4 configuration obtained from GGA+\textit{SO}+\textit{U}.
The Fermi level is set to zero. Note that \textit{s} and \textit{p} states of Mn and \textit{s} states of O are magnified.}
%\end{minipage}
\end{figure}
%%%%%%%%%%%%%%%%%%%%

%%%%%%%%%%%%%%%%%%%  FIG 5 %%%%%%%%%%%%%
\begin{figure*}
%\begin{minipage}{\textwidth}
\hspace{0.10in}
\includegraphics[scale=.5]{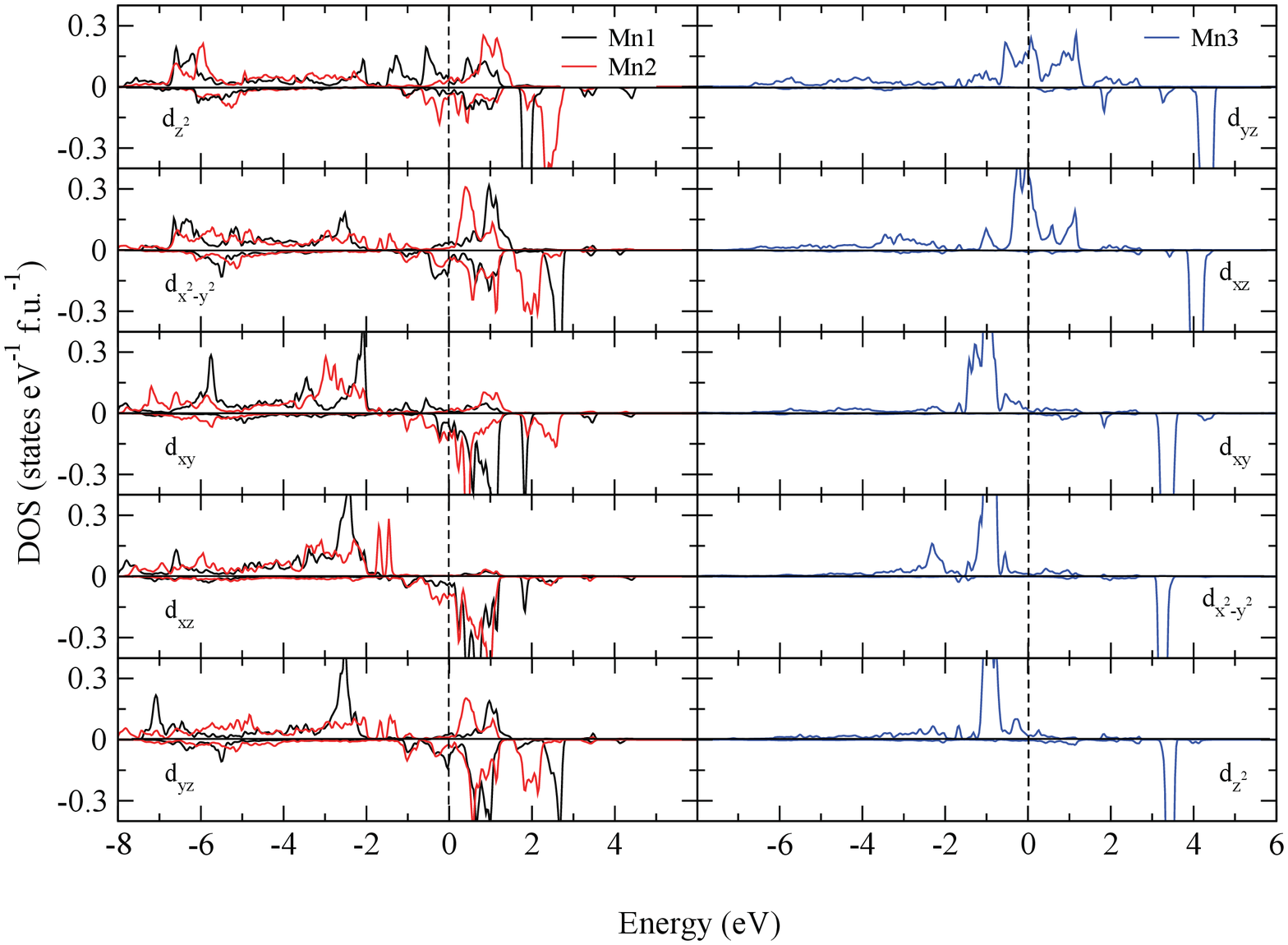}
\caption{\label{fig:oprojdos} (Color online) Orbital (\textit{d}) decomposed DOS for Mn1$^{o}$, Mn2$^{o}$ and Mn3$^{t}$ in Mn$_{5}$O$_{8}$ for the
ground-state M4 configuration. The Fermi level is set to zero.}
%\vspace{0.10in}
\end{figure*}
%%%%%%%%%%%%%%%%%%%%

\par
The Mn \textit{d} levels in a trigonal prismatic coordination split into non-degenerate 1a (d$_{z^{2}}$), doubly degenerate 1e (d$_{xy}$,d$_{x^{2}-y^{2}}$),
and doubly degenerate 2e (d$_{xz}$,d$_{yz}$) levels. As there are five \textit{d} electrons present in the Mn$^{2+}$ ion reside at Mn3 site one would expect that these five electrons occupy the above mentioned levels. For the pure ionic case, the spin moment at the Mn3$^{t}$ site in low-spin(LS; 1a$^{2}$1e$^{3}$2e$^{0}$), intermediate-spin (IS; 1a$^{2}$1e$^{2}$2e$^{1}$), and high-spin (HS; 1a$^{1}$1e$^{2}$2e$^{2}$) configurations will then be 1, 3, and 5 $\mu_{B}$, respectively.
Experimentally measured magnetic moment at the Mn3$^{t}$ site is 3.8 $\mu_{B}$, whereas our calculated magnetic moment gave a somewhat higher spin moment of
4.18 $\mu_{B}$.
The site-projected DOS for Mn and O atoms of Mn$_{5}$O$_{8}$ in the ground state M4 configuration are shown in Fig.~\ref{fig:lprojdos}. The bands originating from oxygen are lying in the energy range -8 and -2 eV and those are almost completely filled. The density of states at E$_{F}$ are mainly contributed by Mn 3\textit{d} states. From this figure, it is clear that the observed metallic behavior in Mn$_{5}$O$_{8}$ is mainly originating from Mn atoms with $\sim$18 $\%$ contribution from oxygen. It may be noted that the minority spin electrons at the fermi level for Mn3 is almost negligible which indicates that the majority spin electrons alone participating in the electrical conductivity, as evident from the Fig.~\ref{fig:lprojdos}. We have observed a noticeable hybridization between the metal Mn\textit{d} and the ligand O\textit{p} states in the whole valence band indicating significant covalency in this system. As a result, the calculated magnetic moments in the Mn sites are smaller than those expected from a pure ionic picture. Consistent with the above view point there is a finite magnetic moment present at the oxygen neighboring to Mn ions.

\par
In order to understand the spin and valence states of Mn1$^{o}$, Mn2$^{o}$, and Mn3$^{t}$ ions in Mn$_{5}$O$_{8}$ we have displayed the orbital-projected \textit{d}-electron DOS of these Mn ions in Fig.~\ref{fig:oprojdos}. This figure reveals that Mn3$^{t}$ is in HS state as it is evident from the occupation of majority spin channel of all five \emph{d} orbitals. For \textit{d}$^{3}$ system, Hund's rule predicted that the electrons will not pair and occupy the t$_{2g}$ orbitals alone by leaving the e$_{g}$ orbitals completely empty. As a result, Fig.~\ref{fig:oprojdos} show almost equal occupation of electrons in all the five \emph{d} orbitals for Mn1$^{o}$ and Mn2$^{o}$. The electron occupation in the e$_{g}$ orbitals of Mn1$^{o}$ and Mn2$^{o}$ bring strong covalent interaction between Mn$^{4+}$ ions and oxygen in the whole valence band region. It is clear from the Fig.~\ref{fig:oprojdos} that higher lying DOS of Mn3 (\textit{d}$_{yz}$ and \textit{d}$_{xz}$) and O1 plays an important role to bring metallic behavior in this compound.

\par
It may be noted that there are two types of exchange interaction one can expect between the magnetic cations in ionic crystals: cation$-$cation and cation-anion-cation (or even cation-anion-anion-cation) interactions. For the case of 90$^{o}$ cation-anion-cation interaction, Goodenough~\cite{goodenough59} have pointed out that if the octahedral interstices of two neighboring cations share a common edge then there is a direct overlap of the \emph{d}$_{xy}$ or \emph{d}$_{yz}$ or \emph{d}$_{xz}$ orbitals of these two cations. As a result, the anion plays a less obvious role in the delocalization-superexchange process. In  Mn$_{5}$O$_{8}$ also Mn ions Mn1O$_{6}$ and Mn2O$_{6}$ are edge shared and hence the interaction between the two neighboring Mn ions can be referred to as cation$-$cation interactions. The cation$-$cation separation is shorter for the oxides than the chlorides and if the t$_{2g}$ orbitals in oxides are half filled then cation$-$cation interactions tend to dominate the 90$^{o}$ superexchange. In the present case also the interatomic distance between Mn1 and Mn2 is smaller (2.81 \AA) and this is comparable to other magnetic oxides such as NaCrO$_{2}$ (2.96 \AA) and NaFeO$_{2}$ (3.02 \AA).~\cite{motida90} It is expected that in the octahedral crystal field the Mn$^{4+}$ ions in the HS state will have a half filled t$_{2g}$ level as we have seen in the orbital decomposed DOS analysis (see Fig.~\ref{fig:oprojdos}). The exchange interaction \emph{J}$^{c-c}_{ij}$ for the two half filled orbitals expected to be negative which results in antiferromagnetic interaction as it was also emphasized by Kanamori.~\cite{kanamori59} In consistent with the above view the Mn$^{4+}$ ions in Mn$_{5}$O$_{8}$ share edges with each other and hence the 90$^{o}$ superexchange interactions is responsible for the observed antiferromagnetic ordering. It may be noted that our experimentally measured Weiss constant $\theta$ value for  Mn$_{5}$O$_{8}$ is -144.5\,K~\cite{gao10} which is indicative for antiferromagnetic superexchange interactions as mentioned above.

\par
When we analyze orbital decomposed DOS for the oxygen \textit{p} states, we found that between $-$2 eV to 0 eV there is no noticeable \textit{p}$_{z}$ states present. However, when we analyze the orbital decomposed \textit{d} states of Mn3 ion we found that majority of \textit{d} states are present within the vicinity of 2 eV from Fermi level. Hence there is no possibility of superexchange path maintains between oxygen \textit{p}$_{z}$ states with Mn3 \textit{d} states as evident from Fig.~\ref{fig:oprojdos} and angular momentum projected DOS(not shown). This could explain why the intralayer coupling between Mn is of ferromagnetic nature.
\par
Mn1 and Mn2 \textit{d} states are spread in the whole valence band from $-$8 eV to Fermi level with dominant contributions from \textit{d}$_{xy}$ and \textit{d}$_{xz}$ states near the Fermi level. In the DOS analysis of Mn1 and Mn2 sites show that there is a strong DOS distribution present between $-$4 eV to $-$2 eV in the valence band and correspondingly the orbital decomposed DOS for oxygen \textit{p} states show that both \textit{p}$_{x}$ and \textit{p}$_{y}$ states are present in this energy range. Hence there is a possibility of superexchange antiferromagnetic interaction between oxygen \textit{p}$_{x}$ and \textit{p}$_{y}$ states with Mn1/Mn2 \textit{d} states between the layer. As a result, one could expect FM interaction within the layer and AFM interaction between the layer resulting \emph{A}-AFM ordering. In consistent with the conclusion arrived from DOS analysis our total energy calculation also show that the \emph{A}-AFM ordering is the lowest energy configuration in this system. In the present observation of \emph{A}-AFM ordering in Mn$_{5}$O$_{8}$ is also supporting the experimentally observed negative Curie temperature value from the susceptibility measurements.~\cite{gao10}

\begin{table}[b]
\caption{Calculated bond strength from Integrated COHP.}
\begin{ruledtabular}
\begin{tabular}{lc}
Interaction & bond strength (eV)\\
\hline
Mn1$-$O1 & -2.18 \\
Mn1$-$O2 & -1.84 \\
Mn2$-$O1 & -2.27 \\
Mn2$-$O2 & -2.17 \\
Mn2$-$O3 & -2.33 \\
Mn3$-$O1 & -1.35 \\
Mn3$-$O2 & -1.27 \\
Mn3$-$O3 & -2.04 \\
Mn1$-$Mn2 & -0.26 \\
Mn2$-$Mn3 &  -0.05
\label{table:bondstrength}
\end{tabular}
\end{ruledtabular}
\end{table}
%%%%%%%%%%%%%%%%%%%%% End Table 1 %%%%%%%%%%%%%%%%%%%%%%%%%%%%%%%%%%

\par
\begin{figure}
\includegraphics[scale=0.45]{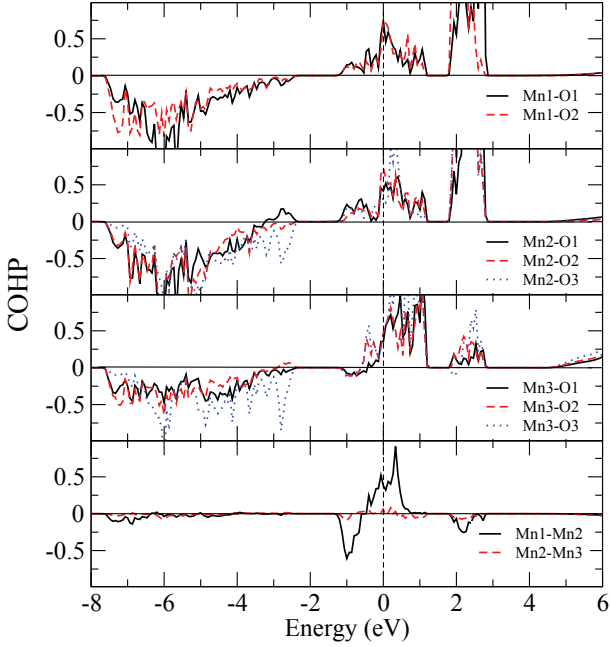}
\caption{\label{fig:cohp} (Color online) The calculated COHP for Mn1$-$O1, Mn1$-$O2, Mn2$-$O1, Mn2$-$O2, Mn2$-$O3, Mn3$-$O1, Mn3$-$O2, Mn3$-$O3, Mn1$-$Mn2 and Mn2$-$Mn3 in ground-state M4 configuration of Mn$_{5}$O$_{8}$}
\end{figure}

The crystal orbital Hamiltonian population (COHP) is calculated using VASP to analyze the bond strength and character of the bonding interactions between the constituents in Mn$_{5}$O$_{8}$.~\cite{deringer11,maintz13} COHP is constructed by weighting the DOS with the corresponding Hamiltonian matrix.~\cite{dronskowski93} This approach provides a qualitative description of the bonding (negative COHP) and antibonding (positive COHP) interaction between atoms. Fig.~\ref{fig:cohp} displays the calculated COHP for the Mn1$-$O1, Mn1$-$O2, Mn2$-$O1, Mn2$-$O2, Mn2$-$O3, Mn3$-$O1, Mn3$-$O2, Mn3$-$O3, Mn1$-$Mn2 and Mn2$-$Mn3 interactions in Mn$_{5}$O$_{8}$. The COHP curves reveal that the Mn$-$O bonding states are present strongly in the region $-$7.5 to $-$2.3 eV of the valence band. In the vicinity of Fermi level (i.e, within -1 eV to E$_{F}$) in Mn$-$O COHP there are noticeable antibonding states present. As the Mn$-$Mn interactions were screened by oxygen ions present between the Mn ions, the calculated Mn$-$Mn bonding states distribution is relatively small. We have calculated the bond strength between various constituents in Mn$_{5}$O$_{8}$ using the Integrated COHP (ICOHP) as we have done earlier.~\cite{ravi2006} The calculated ICOHP for various constituents in Mn$_{5}$O$_{8}$ reflecting the bond strength between the atoms are listed in Table~\ref{table:bondstrength}. According to Table~\ref{table:bondstrength} the strongest bonding interaction is present between Mn$-$O in Mn$_{5}$O$_{8}$. The bonding interaction between Mn2$-$Mn3 is very weak since the bond distance between Mn$^{4+}$ ions within the layer and Mn$^{2+}$ ion reside at the trigonal prismatic coordination is relatively high ($\sim$ 3\AA). The MnO$_{6}$ octahedra are highly distorted and hence the Mn2-O distances are varying between 1.890 \AA\ to 1.955 \AA\ and correspondingly the calculated bond strength is varying $-$2.17 eV to $-$2.33 eV. The calculated bond length for Mn1$-$O$_{a}$ is 1.993 \AA\ and that for Mn2$-$O$_{a}$ is 1.938 \AA\ where, O$_{a}$ is apical oxygen in the octahedra. Hence one can expect that Mn2$-$O$_{a}$ bond will have stronger bond strength than that for Mn1$-$O$_{a}$. Consistent with the above view our calculated ICOHP for Mn2$-$O3$_{a}$ is larger than that for Mn1$-$O2$_{a}$. If there are strong bonds present between magnetic cations with anions, one can expect that the magnetic moment will quench. The calculated bond strength between constituents obtained based on ICOHP values listed in Table~\ref{table:bondstrength} are reflecting the variation in magnetic moments in the corresponding Mn atoms (see Table~\ref{table:moment}). Moreover the difference in the average bond strength for Mn1$-$O and Mn2$-$O indicate that the Mn atoms in Mn$_{5}$O$_{8}$ are in two different oxidation states consistent with our DOS analysis.

%%%%%%%%%%%%%%%%%%%%%%%%% FIG.7 %%%%%%%%%%%%%%%%%%%%%%%%%%%%%%%%%%
%%%%%%%%%%%%%%%%%%%%%%%%%% END %%%%%%%%%%%%%%%%%%%%%%%%%%%%%%%%%%%

\begin{table}
\caption{Selected inter-atomic distances (\AA) and bond valence sum (BVS) for Mn$_{5}$O$_{8}$ at 9 K.
Calculated standard deviations are listed in parentheses.}
\begin{ruledtabular}
\begin{tabular}{lccc}
& Mn1 & Mn2 & Mn3\\
\hline
O1 & 1.91(2)$\times$4 & 1.89(2)$\times$2 & 2.15(2)$\times$4\\
O2 & 1.99(2)$\times$2 &	1.96(2)$\times$2 & 2.10(3) \\
O3 & & 1.94(2)$\times$2 & 2.00(3) \\
\hline
BVS & 3.7(2) & 3.7(2) &	2.3(2)\
\label{table:bvs}
\let\thefootnote\relax\footnote{Note: BVS values were calculated by iteration from initial arbitrary mixed-valence value. The bond valence parameter was assumed by taking fraction of
standard literature values; Mn$^{2+}$$-$O$^{2-}$, Mn$^{3+}$$-$O$^{2-}$, and Mn$^{4+}$$-$O$^{2-}$ are 1.79, 1.76, and 1.753 respectively.}

%\footnote{Note: BVS values were calculated by iteration from initial arbitrary mixed-valence value. The bond valence parameter was assumed by taking fraction of
%standard literature values; Mn$^{2+}$-O$^{2-}$, Mn$^{3+}$-O$^{2-}$, and Mn$^{4+}$-O$^{2-}$ are 1.79, 1.76, and 1.753 respectively.}
\end{tabular}
\end{ruledtabular}
\end{table}
%%%%%%%%%%%%%%%%%%%%% End Table 1 %%%%%%%%%%%%%%%%%%%%%%%%%%%%%%%%%%

\par
The formal oxidation states of the octahedral and trigonal-prismatic Mn atoms can be evaluated from bond-valence-sum (BVS) calculations, by using bond-valence parameters of R$_{ij}$(Mn$^{2+}$) = 1.79, R$_{ij}$(Mn$^{3+}$) = 1.76 and R$_{ij}$(Mn$^{4+}$) = 1.753, and bond distances of d$_{ij}$, yielded from NPD analysis at 9 K, in the formula $\Sigma$exp[(R$_{ij}$-d$_{ij}$)/0.37]. The results indicate that octahedral coordinated Mn atoms at 2$c$ (Mn1) and 4$g$ (Mn2) sites are tetravalent and trigonal-prismatic coordinated Mn atom at 4$i$ (Mn3) site is divalent (see Table~\ref{table:bvs}), even though the BVS values for octahedral and trigonal-prismatic polyhedral deviate slightly from the ideal value of 3.7(2) and 2.3(2), respectively.

\par
The valence states of Mn in Mn$_{5}$O$_{8}$ can be written as Mn$_{2}^{2+}$Mn$_{3}^{4+}$O$_{8}$, the spin-only magnetic moment for Mn$_{5}$O$_{8}$ can be calculated from the expression

\begin{equation}
\mu_{\rm eff} = \sqrt{\frac{2}{5}[\ \mu_{\rm Mn(II)}]^{2} + \frac{3}{5}[\ \mu_{\rm Mn(IV)}]^{2}}
\end{equation}

An effective magnetic moment measured from magnetic susceptibility study is $\mu_{\rm eff} = 4.803\mu_{B}$. The calculated effective magnetic moment using $\mu_{\rm Mn(II)}$ = $5.9\mu_{B}$ and $\mu_{\rm Mn(IV)}$ = $3.87\mu_{B}$ is $\mu_{\rm eff} = 4.786\mu_{B}$. The close agreement between the calculated spin-only magnetic moment and the experimental moment confirms the presence of mixed valence states in Mn$_{5}$O$_{8}$. As the experimentally measured effective magnetic moment is near their spin-only value calculated theoretically, it is apparent that the orbital angular moment is quenched due to the ligand field splitting which was also evident from Table~\ref{table:orbmoment} that our calculated orbital moment is very small and only the spin moment is mainly contributing to the total magnetic moment.
\par
In previous study,~\cite{gao10} we have mentioned the possibility of Mn$_{3}$O$_{4}$~\cite{seo04} as secondary phase in master Mn$_{5}$O$_{8}$ sample, based on the observed magnetic ferroic transition around 40\,K. In present study, we have measured NPD at 70\,K (not shown), i.e. above 40\,K but still below T$_{N}$=133\,K. However we are unable to notice any unique changes comparing to the NPD data at 9\,K. Therefore, we conclude that there is no major contribution from secondary phase of Mn$_{3}$O$_{4}$ in collected NPD data.

\section{Summary}
To summarize, magnetic ordering and the mixed valent nature of Mn$_{5}$O$_{8}$ has been reported using fully relativistic density functional theory calculations and NPD measurements. From our total energy calculation we have predicted  that Mn$_{5}$O$_{8}$ orders antiferromagnetically with the spin orientation along [100] direction in accordance with our NPD measurements. Among the four different magnetic models proposed from the experiments, M4 magnetic configuration is found to be the ground state magnetic structure in Mn$_{5}$O$_{8}$. Density of states calculation shows a large DOS at E$_{F}$ in the ground state magnetic structure showing the metallic nature of Mn$_{5}$O$_{8}$. Calculations without the inclusion of Hubbard \textit{U} leads to ferrimagnetic half metal as ground state contradictory to experimental findings, indicating the presence of strong Coulomb correlation effect in this material. However we include Coulomb correlation effect in to the calculations correctly predict antiferromagnetic metal as a ground state. Presence of two different magnetic moments along with the different density of states distribution in the Mn sites so the mixed valent behavior of Mn ions. Bond strength analysis based on COHP between constituents indicating strong anisotropy in the bonding behavior which results in layered nature of its crystal structure. The orbital decomposed DOS analysis of Mn \emph{d} states show that Mn ions in the Mn$^{4+}$ and Mn$^{2+}$ oxidation states with high spin magnetic configuration. Our BVS analysis also support the presence of two different oxidation states such as 4+ and 2+ in this compound. The calculated orbital moment using relativistic spin polarized calculation with orbital polarization correction yielded very small orbital moment and hence the spin only moment is sufficient to account the experimentally measured effective moment from susceptibility measurements. The over all results indicate that Mn$_{5}$O$_{8}$ can be described as a strongly correlated mixed valent antiferromagnetic metal.

\section{Acknowledgement}
The authors are grateful to the Research Council of Norway for computing time on the Norwegian supercomputer facilities. This research was supported by the Indo-Norwegian Cooperative Program (INCP) via Grant No. F.No. 58-12/2014(IC).

\appendix*
\section{Group theory analysis for Mn$_{5}$O$_{8}$}
To determine magnetic structure of Mn$_{5}$O$_{8}$, the possible magnetic structures, which are compatible with the crystal symmetry, were investigated by the method described by Bertaut~\cite{bertaut68}. The magnetic structures are defined by the basis vectors of the irreducible representations of the little group $\Gamma$$_{k}$ which for a propagation vector k = (-1, 0, 1) coincides with the crystallographic space group.
\par
In the case of Mn$_{5}$O$_{8}$ the Mn atoms are in wyckoff sites at 2c, 4g, and 4i under space group of C2/m. The decomposition of the representation $\Gamma$, describing the transformation properties of the magnetic moments is given by $\Gamma$ = $\Gamma$$_{1}$ + 2$\Gamma$$_{3}$ in Kovalev's notation.~\cite{kovalev85} The solutions defined by $\Gamma$$_{1}$ correspond to FM ordering within layers and AFM ordering between each layer with magnetic moments parallel to the \textit{b} direction (M3 in Fig. 2). $\Gamma$$_{5}$ defines similar ordering with F1 in the \textit{ac} plane (M4 in Fig. 2).

\bibliography{Mn5O8}		% expects file "ref.bib"

\begin{thebibliography}{62}%
\makeatletter
\providecommand \@ifxundefined [1]{%
 \@ifx{#1\undefined}
}%
\providecommand \@ifnum [1]{%
 \ifnum #1\expandafter \@firstoftwo
 \else \expandafter \@secondoftwo
 \fi
}%
\providecommand \@ifx [1]{%
 \ifx #1\expandafter \@firstoftwo
 \else \expandafter \@secondoftwo
 \fi
}%
\providecommand \natexlab [1]{#1}%
\providecommand \enquote  [1]{``#1''}%
\providecommand \bibnamefont  [1]{#1}%
\providecommand \bibfnamefont [1]{#1}%
\providecommand \citenamefont [1]{#1}%
\providecommand \href@noop [0]{\@secondoftwo}%
\providecommand \href [0]{\begingroup \@sanitize@url \@href}%
\providecommand \@href[1]{\@@startlink{#1}\@@href}%
\providecommand \@@href[1]{\endgroup#1\@@endlink}%
\providecommand \@sanitize@url [0]{\catcode `\\12\catcode `\$12\catcode
  `\&12\catcode `\#12\catcode `\^12\catcode `\_12\catcode `\%12\relax}%
\providecommand \@@startlink[1]{}%
\providecommand \@@endlink[0]{}%
\providecommand \url  [0]{\begingroup\@sanitize@url \@url }%
\providecommand \@url [1]{\endgroup\@href {#1}{\urlprefix }}%
\providecommand \urlprefix  [0]{URL }%
\providecommand \Eprint [0]{\href }%
\providecommand \doibase [0]{http://dx.doi.org/}%
\providecommand \selectlanguage [0]{\@gobble}%
\providecommand \bibinfo  [0]{\@secondoftwo}%
\providecommand \bibfield  [0]{\@secondoftwo}%
\providecommand \translation [1]{[#1]}%
\providecommand \BibitemOpen [0]{}%
\providecommand \bibitemStop [0]{}%
\providecommand \bibitemNoStop [0]{.\EOS\space}%
\providecommand \EOS [0]{\spacefactor3000\relax}%
\providecommand \BibitemShut  [1]{\csname bibitem#1\endcsname}%
\let\auto@bib@innerbib\@empty
%</preamble>
\bibitem [{\citenamefont {Van~Elp}\ \emph {et~al.}(1991)\citenamefont
  {Van~Elp}, \citenamefont {Potze}, \citenamefont {Eskes}, \citenamefont
  {Berger},\ and\ \citenamefont {Sawatzky}}]{van91}%
  \BibitemOpen
  \bibfield  {author} {\bibinfo {author} {\bibfnamefont {J.}~\bibnamefont
  {Van~Elp}}, \bibinfo {author} {\bibfnamefont {R.}~\bibnamefont {Potze}},
  \bibinfo {author} {\bibfnamefont {H.}~\bibnamefont {Eskes}}, \bibinfo
  {author} {\bibfnamefont {R.}~\bibnamefont {Berger}}, \ and\ \bibinfo {author}
  {\bibfnamefont {G.}~\bibnamefont {Sawatzky}},\ }\href@noop {} {\bibfield
  {journal} {\bibinfo  {journal} {Phys. Rev. B}\ }\textbf {\bibinfo {volume}
  {44}},\ \bibinfo {pages} {1530} (\bibinfo {year} {1991})}\BibitemShut
  {NoStop}%
\bibitem [{\citenamefont {Geller}(1971)}]{geller71}%
  \BibitemOpen
  \bibfield  {author} {\bibinfo {author} {\bibfnamefont {S.}~\bibnamefont
  {Geller}},\ }\href@noop {} {\bibfield  {journal} {\bibinfo  {journal} {Acta.
  Crystallogr. Sect. B}\ }\textbf {\bibinfo {volume} {27}},\ \bibinfo {pages}
  {821} (\bibinfo {year} {1971})}\BibitemShut {NoStop}%
\bibitem [{\citenamefont {Gao}\ \emph {et~al.}(2009{\natexlab{a}})\citenamefont
  {Gao}, \citenamefont {Fjellv{\aa}g},\ and\ \citenamefont {Norby}}]{gao09}%
  \BibitemOpen
  \bibfield  {author} {\bibinfo {author} {\bibfnamefont {T.}~\bibnamefont
  {Gao}}, \bibinfo {author} {\bibfnamefont {H.}~\bibnamefont {Fjellv{\aa}g}}, \
  and\ \bibinfo {author} {\bibfnamefont {P.}~\bibnamefont {Norby}},\
  }\href@noop {} {\bibfield  {journal} {\bibinfo  {journal} {Analytica chimica
  acta}\ }\textbf {\bibinfo {volume} {648}},\ \bibinfo {pages} {235} (\bibinfo
  {year} {2009}{\natexlab{a}})}\BibitemShut {NoStop}%
\bibitem [{\citenamefont {Franchini}\ \emph {et~al.}(2007)\citenamefont
  {Franchini}, \citenamefont {Podloucky}, \citenamefont {Paier}, \citenamefont
  {Marsman},\ and\ \citenamefont {Kresse}}]{franchi07}%
  \BibitemOpen
  \bibfield  {author} {\bibinfo {author} {\bibfnamefont {C.}~\bibnamefont
  {Franchini}}, \bibinfo {author} {\bibfnamefont {R.}~\bibnamefont
  {Podloucky}}, \bibinfo {author} {\bibfnamefont {J.}~\bibnamefont {Paier}},
  \bibinfo {author} {\bibfnamefont {M.}~\bibnamefont {Marsman}}, \ and\
  \bibinfo {author} {\bibfnamefont {G.}~\bibnamefont {Kresse}},\ }\href@noop {}
  {\bibfield  {journal} {\bibinfo  {journal} {Phys. Rev. B}\ }\textbf {\bibinfo
  {volume} {75}},\ \bibinfo {pages} {195128} (\bibinfo {year}
  {2007})}\BibitemShut {NoStop}%
\bibitem [{\citenamefont {Ravindran}\ \emph {et~al.}(2002)\citenamefont
  {Ravindran}, \citenamefont {Kjekshus}, \citenamefont {Fjellv{\aa}g},
  \citenamefont {Delin},\ and\ \citenamefont {Eriksson}}]{ravi02}%
  \BibitemOpen
  \bibfield  {author} {\bibinfo {author} {\bibfnamefont {P.}~\bibnamefont
  {Ravindran}}, \bibinfo {author} {\bibfnamefont {A.}~\bibnamefont {Kjekshus}},
  \bibinfo {author} {\bibfnamefont {H.}~\bibnamefont {Fjellv{\aa}g}}, \bibinfo
  {author} {\bibfnamefont {A.}~\bibnamefont {Delin}}, \ and\ \bibinfo {author}
  {\bibfnamefont {O.}~\bibnamefont {Eriksson}},\ }\href@noop {} {\bibfield
  {journal} {\bibinfo  {journal} {Phys. Rev. B}\ }\textbf {\bibinfo {volume}
  {65}},\ \bibinfo {pages} {064445} (\bibinfo {year} {2002})}\BibitemShut
  {NoStop}%
\bibitem [{\citenamefont {Shannon}(1976)}]{shannon76}%
  \BibitemOpen
  \bibfield  {author} {\bibinfo {author} {\bibfnamefont {R.~D.}\ \bibnamefont
  {Shannon}},\ }\href@noop {} {\bibfield  {journal} {\bibinfo  {journal} {Acta
  Crystallogr. Sec. A}\ }\textbf {\bibinfo {volume} {32}},\ \bibinfo {pages}
  {751} (\bibinfo {year} {1976})}\BibitemShut {NoStop}%
\bibitem [{\citenamefont {Thackeray}(1997)}]{thackeray97}%
  \BibitemOpen
  \bibfield  {author} {\bibinfo {author} {\bibfnamefont {M.~M.}\ \bibnamefont
  {Thackeray}},\ }\href@noop {} {\bibfield  {journal} {\bibinfo  {journal}
  {Progress in Solid State Chemistry}\ }\textbf {\bibinfo {volume} {25}},\
  \bibinfo {pages} {1} (\bibinfo {year} {1997})}\BibitemShut {NoStop}%
\bibitem [{\citenamefont {Post}(1999)}]{jeffrey99}%
  \BibitemOpen
  \bibfield  {author} {\bibinfo {author} {\bibfnamefont {J.~E.}\ \bibnamefont
  {Post}},\ }\href@noop {} {\bibfield  {journal} {\bibinfo  {journal} {Proc
  Natl Acad Sci}\ }\textbf {\bibinfo {volume} {96}},\ \bibinfo {pages} {3447}
  (\bibinfo {year} {1999})}\BibitemShut {NoStop}%
\bibitem [{\citenamefont {Vidya}\ \emph {et~al.}(2004)\citenamefont {Vidya},
  \citenamefont {Ravindran}, \citenamefont {Vajeeston}, \citenamefont
  {Kjekshus},\ and\ \citenamefont {Fjellv{\aa}g}}]{vidya04}%
  \BibitemOpen
  \bibfield  {author} {\bibinfo {author} {\bibfnamefont {R.}~\bibnamefont
  {Vidya}}, \bibinfo {author} {\bibfnamefont {P.}~\bibnamefont {Ravindran}},
  \bibinfo {author} {\bibfnamefont {P.}~\bibnamefont {Vajeeston}}, \bibinfo
  {author} {\bibfnamefont {A.}~\bibnamefont {Kjekshus}}, \ and\ \bibinfo
  {author} {\bibfnamefont {H.}~\bibnamefont {Fjellv{\aa}g}},\ }\href@noop {}
  {\bibfield  {journal} {\bibinfo  {journal} {Phys. Rev. B}\ }\textbf {\bibinfo
  {volume} {69}},\ \bibinfo {pages} {092405} (\bibinfo {year}
  {2004})}\BibitemShut {NoStop}%
\bibitem [{\citenamefont {Vidya}\ \emph {et~al.}(2002)\citenamefont {Vidya},
  \citenamefont {Ravindran}, \citenamefont {Kjekshus},\ and\ \citenamefont
  {Fjellv{\aa}g}}]{vidya02}%
  \BibitemOpen
  \bibfield  {author} {\bibinfo {author} {\bibfnamefont {R.}~\bibnamefont
  {Vidya}}, \bibinfo {author} {\bibfnamefont {P.}~\bibnamefont {Ravindran}},
  \bibinfo {author} {\bibfnamefont {A.}~\bibnamefont {Kjekshus}}, \ and\
  \bibinfo {author} {\bibfnamefont {H.}~\bibnamefont {Fjellv{\aa}g}},\
  }\href@noop {} {\bibfield  {journal} {\bibinfo  {journal} {Phys. Rev. B}\
  }\textbf {\bibinfo {volume} {65}},\ \bibinfo {pages} {144422} (\bibinfo
  {year} {2002})}\BibitemShut {NoStop}%
\bibitem [{\citenamefont {Franchini}\ \emph {et~al.}(2005)\citenamefont
  {Franchini}, \citenamefont {Bayer}, \citenamefont {Podloucky}, \citenamefont
  {Paier},\ and\ \citenamefont {Kresse}}]{franchi05}%
  \BibitemOpen
  \bibfield  {author} {\bibinfo {author} {\bibfnamefont {C.}~\bibnamefont
  {Franchini}}, \bibinfo {author} {\bibfnamefont {V.}~\bibnamefont {Bayer}},
  \bibinfo {author} {\bibfnamefont {R.}~\bibnamefont {Podloucky}}, \bibinfo
  {author} {\bibfnamefont {J.}~\bibnamefont {Paier}}, \ and\ \bibinfo {author}
  {\bibfnamefont {G.}~\bibnamefont {Kresse}},\ }\href@noop {} {\bibfield
  {journal} {\bibinfo  {journal} {Phys. Rev. B}\ }\textbf {\bibinfo {volume}
  {72}},\ \bibinfo {pages} {045132} (\bibinfo {year} {2005})}\BibitemShut
  {NoStop}%
\bibitem [{\citenamefont {Imada}\ \emph {et~al.}(1998)\citenamefont {Imada},
  \citenamefont {Fujimori},\ and\ \citenamefont {Tokura}}]{imada98}%
  \BibitemOpen
  \bibfield  {author} {\bibinfo {author} {\bibfnamefont {M.}~\bibnamefont
  {Imada}}, \bibinfo {author} {\bibfnamefont {A.}~\bibnamefont {Fujimori}}, \
  and\ \bibinfo {author} {\bibfnamefont {Y.}~\bibnamefont {Tokura}},\
  }\href@noop {} {\bibfield  {journal} {\bibinfo  {journal} {Rev. Mod. Phys.}\
  }\textbf {\bibinfo {volume} {70}},\ \bibinfo {pages} {1039} (\bibinfo {year}
  {1998})}\BibitemShut {NoStop}%
\bibitem [{\citenamefont {Mott}(1968)}]{mott68}%
  \BibitemOpen
  \bibfield  {author} {\bibinfo {author} {\bibfnamefont {N.}~\bibnamefont
  {Mott}},\ }\href@noop {} {\bibfield  {journal} {\bibinfo  {journal} {Rev.
  Mod. Phys.}\ }\textbf {\bibinfo {volume} {40}},\ \bibinfo {pages} {677}
  (\bibinfo {year} {1968})}\BibitemShut {NoStop}%
\bibitem [{\citenamefont {Harrison}(2008)}]{harri08}%
  \BibitemOpen
  \bibfield  {author} {\bibinfo {author} {\bibfnamefont {W.~A.}\ \bibnamefont
  {Harrison}},\ }\href@noop {} {\bibfield  {journal} {\bibinfo  {journal}
  {Phys. Rev. B}\ }\textbf {\bibinfo {volume} {77}},\ \bibinfo {pages} {245103}
  (\bibinfo {year} {2008})}\BibitemShut {NoStop}%
\bibitem [{\citenamefont {Shaked}\ \emph {et~al.}(1988)\citenamefont {Shaked},
  \citenamefont {Faber~Jr},\ and\ \citenamefont {Hitterman}}]{shaked88}%
  \BibitemOpen
  \bibfield  {author} {\bibinfo {author} {\bibfnamefont {H.}~\bibnamefont
  {Shaked}}, \bibinfo {author} {\bibfnamefont {J.}~\bibnamefont {Faber~Jr}}, \
  and\ \bibinfo {author} {\bibfnamefont {R.}~\bibnamefont {Hitterman}},\
  }\href@noop {} {\bibfield  {journal} {\bibinfo  {journal} {Phys. Rev. B}\
  }\textbf {\bibinfo {volume} {38}},\ \bibinfo {pages} {11901} (\bibinfo {year}
  {1988})}\BibitemShut {NoStop}%
\bibitem [{\citenamefont {Dwight}\ and\ \citenamefont
  {Menyuk}(1960)}]{dwight60}%
  \BibitemOpen
  \bibfield  {author} {\bibinfo {author} {\bibfnamefont {K.}~\bibnamefont
  {Dwight}}\ and\ \bibinfo {author} {\bibfnamefont {N.}~\bibnamefont
  {Menyuk}},\ }\href@noop {} {\bibfield  {journal} {\bibinfo  {journal} {Phys.
  Rev. B}\ }\textbf {\bibinfo {volume} {199}},\ \bibinfo {pages} {1470}
  (\bibinfo {year} {1960})}\BibitemShut {NoStop}%
\bibitem [{\citenamefont {Jensen}\ and\ \citenamefont
  {Nielsen}(1974)}]{jensen74}%
  \BibitemOpen
  \bibfield  {author} {\bibinfo {author} {\bibfnamefont {G.}~\bibnamefont
  {Jensen}}\ and\ \bibinfo {author} {\bibfnamefont {O.}~\bibnamefont
  {Nielsen}},\ }\href@noop {} {\bibfield  {journal} {\bibinfo  {journal} {J.
  Phys. C}\ }\textbf {\bibinfo {volume} {7}},\ \bibinfo {pages} {409} (\bibinfo
  {year} {1974})}\BibitemShut {NoStop}%
\bibitem [{\citenamefont {Srinivasan}\ and\ \citenamefont
  {Seehra}(1983)}]{srini83}%
  \BibitemOpen
  \bibfield  {author} {\bibinfo {author} {\bibfnamefont {G.}~\bibnamefont
  {Srinivasan}}\ and\ \bibinfo {author} {\bibfnamefont {M.~S.}\ \bibnamefont
  {Seehra}},\ }\href@noop {} {\bibfield  {journal} {\bibinfo  {journal} {Phys.
  Rev. B}\ }\textbf {\bibinfo {volume} {28}},\ \bibinfo {pages} {1} (\bibinfo
  {year} {1983})}\BibitemShut {NoStop}%
\bibitem [{\citenamefont {Yoshimori}(1959)}]{yoshi59}%
  \BibitemOpen
  \bibfield  {author} {\bibinfo {author} {\bibfnamefont {A.}~\bibnamefont
  {Yoshimori}},\ }\href@noop {} {\bibfield  {journal} {\bibinfo  {journal} {J.
  Phys. Soc. Jpn.}\ }\textbf {\bibinfo {volume} {14}},\ \bibinfo {pages} {807}
  (\bibinfo {year} {1959})}\BibitemShut {NoStop}%
\bibitem [{\citenamefont {Sato}\ \emph {et~al.}(2001)\citenamefont {Sato},
  \citenamefont {Wakiya}, \citenamefont {Enoki}, \citenamefont {Kiyama},
  \citenamefont {Wakabayashi}, \citenamefont {Nakao},\ and\ \citenamefont
  {Murakami}}]{sato01}%
  \BibitemOpen
  \bibfield  {author} {\bibinfo {author} {\bibfnamefont {H.}~\bibnamefont
  {Sato}}, \bibinfo {author} {\bibfnamefont {K.}~\bibnamefont {Wakiya}},
  \bibinfo {author} {\bibfnamefont {T.}~\bibnamefont {Enoki}}, \bibinfo
  {author} {\bibfnamefont {T.}~\bibnamefont {Kiyama}}, \bibinfo {author}
  {\bibfnamefont {Y.}~\bibnamefont {Wakabayashi}}, \bibinfo {author}
  {\bibfnamefont {H.}~\bibnamefont {Nakao}}, \ and\ \bibinfo {author}
  {\bibfnamefont {Y.}~\bibnamefont {Murakami}},\ }\href@noop {} {\bibfield
  {journal} {\bibinfo  {journal} {J. Phys. Soc. Jpn.}\ }\textbf {\bibinfo
  {volume} {70}},\ \bibinfo {pages} {37} (\bibinfo {year} {2001})}\BibitemShut
  {NoStop}%
\bibitem [{\citenamefont {Sugawara}\ \emph {et~al.}(1991)\citenamefont
  {Sugawara}, \citenamefont {Ohno},\ and\ \citenamefont {Matsuki}}]{suga91}%
  \BibitemOpen
  \bibfield  {author} {\bibinfo {author} {\bibfnamefont {M.}~\bibnamefont
  {Sugawara}}, \bibinfo {author} {\bibfnamefont {M.}~\bibnamefont {Ohno}}, \
  and\ \bibinfo {author} {\bibfnamefont {K.}~\bibnamefont {Matsuki}},\
  }\href@noop {} {\bibfield  {journal} {\bibinfo  {journal} {Chemistry
  Letters}\ ,\ \bibinfo {pages} {1465}} (\bibinfo {year} {1991})}\BibitemShut
  {NoStop}%
\bibitem [{\citenamefont {Azzoni}\ \emph {et~al.}(1999)\citenamefont {Azzoni},
  \citenamefont {Mozzati}, \citenamefont {Galinetto}, \citenamefont {Paleari},
  \citenamefont {Massarotti}, \citenamefont {Capsoni},\ and\ \citenamefont
  {Bini}}]{azzoni99}%
  \BibitemOpen
  \bibfield  {author} {\bibinfo {author} {\bibfnamefont {C.}~\bibnamefont
  {Azzoni}}, \bibinfo {author} {\bibfnamefont {M.}~\bibnamefont {Mozzati}},
  \bibinfo {author} {\bibfnamefont {P.}~\bibnamefont {Galinetto}}, \bibinfo
  {author} {\bibfnamefont {A.}~\bibnamefont {Paleari}}, \bibinfo {author}
  {\bibfnamefont {V.}~\bibnamefont {Massarotti}}, \bibinfo {author}
  {\bibfnamefont {D.}~\bibnamefont {Capsoni}}, \ and\ \bibinfo {author}
  {\bibfnamefont {M.}~\bibnamefont {Bini}},\ }\href@noop {} {\bibfield
  {journal} {\bibinfo  {journal} {Solid State Commun.}\ }\textbf {\bibinfo
  {volume} {112}},\ \bibinfo {pages} {375} (\bibinfo {year}
  {1999})}\BibitemShut {NoStop}%
\bibitem [{\citenamefont {Rask}\ and\ \citenamefont {Buseck}(1986)}]{rask86}%
  \BibitemOpen
  \bibfield  {author} {\bibinfo {author} {\bibfnamefont {J.~H.}\ \bibnamefont
  {Rask}}\ and\ \bibinfo {author} {\bibfnamefont {P.~R.}\ \bibnamefont
  {Buseck}},\ }\href@noop {} {\bibfield  {journal} {\bibinfo  {journal}
  {American Mineralogist}\ }\textbf {\bibinfo {volume} {71}},\ \bibinfo {pages}
  {805} (\bibinfo {year} {1986})}\BibitemShut {NoStop}%
\bibitem [{\citenamefont {Auerbach}\ \emph {et~al.}(2004)\citenamefont
  {Auerbach}, \citenamefont {Carrado},\ and\ \citenamefont
  {Dutta}}]{auerbach04}%
  \BibitemOpen
  \bibfield  {author} {\bibinfo {author} {\bibfnamefont {S.~M.}\ \bibnamefont
  {Auerbach}}, \bibinfo {author} {\bibfnamefont {K.~A.}\ \bibnamefont
  {Carrado}}, \ and\ \bibinfo {author} {\bibfnamefont {P.~K.}\ \bibnamefont
  {Dutta}},\ }\href@noop {} {\emph {\bibinfo {title} {Handbook of Layered
  Materials}}}\ (\bibinfo  {publisher} {Marcel Dekker, Inc.},\ \bibinfo
  {address} {New York},\ \bibinfo {year} {2004})\ Chap.~\bibinfo {chapter} {9},
  pp.\ \bibinfo {pages} {475--508}\BibitemShut {NoStop}%
\bibitem [{\citenamefont {Oswald}\ and\ \citenamefont
  {Wampetich}(1967)}]{oswald67}%
  \BibitemOpen
  \bibfield  {author} {\bibinfo {author} {\bibfnamefont {H.~R.}\ \bibnamefont
  {Oswald}}\ and\ \bibinfo {author} {\bibfnamefont {M.~J.}\ \bibnamefont
  {Wampetich}},\ }\href@noop {} {\bibfield  {journal} {\bibinfo  {journal}
  {Helv. Chim. Acta}\ }\textbf {\bibinfo {volume} {50}},\ \bibinfo {pages}
  {2023} (\bibinfo {year} {1967})}\BibitemShut {NoStop}%
\bibitem [{\citenamefont {Yamamoto}\ \emph {et~al.}(1973)\citenamefont
  {Yamamoto}, \citenamefont {Kiyama},\ and\ \citenamefont {Takada}}]{yama73}%
  \BibitemOpen
  \bibfield  {author} {\bibinfo {author} {\bibfnamefont {N.}~\bibnamefont
  {Yamamoto}}, \bibinfo {author} {\bibfnamefont {M.}~\bibnamefont {Kiyama}}, \
  and\ \bibinfo {author} {\bibfnamefont {T.}~\bibnamefont {Takada}},\
  }\href@noop {} {\bibfield  {journal} {\bibinfo  {journal} {J. Appl. Phys.}\
  }\textbf {\bibinfo {volume} {12}},\ \bibinfo {pages} {1827} (\bibinfo {year}
  {1973})}\BibitemShut {NoStop}%
\bibitem [{\citenamefont {Punnoose}\ \emph {et~al.}(2001)\citenamefont
  {Punnoose}, \citenamefont {Magnone},\ and\ \citenamefont
  {Seehra}}]{punnoose01}%
  \BibitemOpen
  \bibfield  {author} {\bibinfo {author} {\bibfnamefont {A.}~\bibnamefont
  {Punnoose}}, \bibinfo {author} {\bibfnamefont {H.}~\bibnamefont {Magnone}}, \
  and\ \bibinfo {author} {\bibfnamefont {M.}~\bibnamefont {Seehra}},\
  }\href@noop {} {\bibfield  {journal} {\bibinfo  {journal} {IEEE Trans.
  Magn.}\ }\textbf {\bibinfo {volume} {37}},\ \bibinfo {pages} {2150} (\bibinfo
  {year} {2001})}\BibitemShut {NoStop}%
\bibitem [{\citenamefont {Gao}\ \emph {et~al.}(2010)\citenamefont {Gao},
  \citenamefont {Norby}, \citenamefont {Krumeich}, \citenamefont {Okamoto},
  \citenamefont {Nesper},\ and\ \citenamefont {Fjellv{\aa}g}}]{gao10}%
  \BibitemOpen
  \bibfield  {author} {\bibinfo {author} {\bibfnamefont {T.}~\bibnamefont
  {Gao}}, \bibinfo {author} {\bibfnamefont {P.}~\bibnamefont {Norby}}, \bibinfo
  {author} {\bibfnamefont {F.}~\bibnamefont {Krumeich}}, \bibinfo {author}
  {\bibfnamefont {H.}~\bibnamefont {Okamoto}}, \bibinfo {author} {\bibfnamefont
  {R.}~\bibnamefont {Nesper}}, \ and\ \bibinfo {author} {\bibfnamefont
  {H.}~\bibnamefont {Fjellv{\aa}g}},\ }\href@noop {} {\bibfield  {journal}
  {\bibinfo  {journal} {J. Phys. Chem. C}\ }\textbf {\bibinfo {volume} {114}},\
  \bibinfo {pages} {922} (\bibinfo {year} {2010})}\BibitemShut {NoStop}%
\bibitem [{\citenamefont {Thota}\ \emph {et~al.}(2010)\citenamefont {Thota},
  \citenamefont {Prasad},\ and\ \citenamefont {Kumar}}]{thota10}%
  \BibitemOpen
  \bibfield  {author} {\bibinfo {author} {\bibfnamefont {S.}~\bibnamefont
  {Thota}}, \bibinfo {author} {\bibfnamefont {B.}~\bibnamefont {Prasad}}, \
  and\ \bibinfo {author} {\bibfnamefont {J.}~\bibnamefont {Kumar}},\
  }\href@noop {} {\bibfield  {journal} {\bibinfo  {journal} {Mater. Sci. Eng.
  B}\ }\textbf {\bibinfo {volume} {167}},\ \bibinfo {pages} {153} (\bibinfo
  {year} {2010})}\BibitemShut {NoStop}%
\bibitem [{\citenamefont {Uddin}\ \emph {et~al.}(2013)\citenamefont {Uddin},
  \citenamefont {Poddar},\ and\ \citenamefont {Ahmad}}]{uddin13}%
  \BibitemOpen
  \bibfield  {author} {\bibinfo {author} {\bibfnamefont {I.}~\bibnamefont
  {Uddin}}, \bibinfo {author} {\bibfnamefont {P.}~\bibnamefont {Poddar}}, \
  and\ \bibinfo {author} {\bibfnamefont {A.}~\bibnamefont {Ahmad}},\
  }\href@noop {} {\bibfield  {journal} {\bibinfo  {journal} {Journal of
  Nanoengineering and Nanomanufacturing}\ }\textbf {\bibinfo {volume} {3}},\
  \bibinfo {pages} {91} (\bibinfo {year} {2013})}\BibitemShut {NoStop}%
\bibitem [{\citenamefont {Zheng}\ \emph {et~al.}(2005)\citenamefont {Zheng},
  \citenamefont {Xu}, \citenamefont {Nishikubo}, \citenamefont {Nishiyama},
  \citenamefont {Higemoto}, \citenamefont {Moon}, \citenamefont {Tanaka},\ and\
  \citenamefont {Otabe}}]{zheng05}%
  \BibitemOpen
  \bibfield  {author} {\bibinfo {author} {\bibfnamefont {X.}~\bibnamefont
  {Zheng}}, \bibinfo {author} {\bibfnamefont {C.}~\bibnamefont {Xu}}, \bibinfo
  {author} {\bibfnamefont {K.}~\bibnamefont {Nishikubo}}, \bibinfo {author}
  {\bibfnamefont {K.}~\bibnamefont {Nishiyama}}, \bibinfo {author}
  {\bibfnamefont {W.}~\bibnamefont {Higemoto}}, \bibinfo {author}
  {\bibfnamefont {W.}~\bibnamefont {Moon}}, \bibinfo {author} {\bibfnamefont
  {E.}~\bibnamefont {Tanaka}}, \ and\ \bibinfo {author} {\bibfnamefont {E.~S.}\
  \bibnamefont {Otabe}},\ }\href@noop {} {\bibfield  {journal} {\bibinfo
  {journal} {Phys. Rev. B}\ }\textbf {\bibinfo {volume} {72}},\ \bibinfo
  {pages} {014464} (\bibinfo {year} {2005})}\BibitemShut {NoStop}%
\bibitem [{\citenamefont {He}\ \emph {et~al.}(2007)\citenamefont {He},
  \citenamefont {Chen}, \citenamefont {Wang}, \citenamefont {Zhou},\ and\
  \citenamefont {Guo}}]{lin07}%
  \BibitemOpen
  \bibfield  {author} {\bibinfo {author} {\bibfnamefont {L.}~\bibnamefont
  {He}}, \bibinfo {author} {\bibfnamefont {C.}~\bibnamefont {Chen}}, \bibinfo
  {author} {\bibfnamefont {N.}~\bibnamefont {Wang}}, \bibinfo {author}
  {\bibfnamefont {W.}~\bibnamefont {Zhou}}, \ and\ \bibinfo {author}
  {\bibfnamefont {L.}~\bibnamefont {Guo}},\ }\href@noop {} {\bibfield
  {journal} {\bibinfo  {journal} {J. Appl. Phys.}\ }\textbf {\bibinfo {volume}
  {102}},\ \bibinfo {pages} {103911} (\bibinfo {year} {2007})}\BibitemShut
  {NoStop}%
\bibitem [{\citenamefont {Han}\ \emph {et~al.}(2006)\citenamefont {Han},
  \citenamefont {Chen}, \citenamefont {Zhong}, \citenamefont {Ramesh},
  \citenamefont {Chen},\ and\ \citenamefont {Widjaja}}]{han06}%
  \BibitemOpen
  \bibfield  {author} {\bibinfo {author} {\bibfnamefont {Y.-F.}\ \bibnamefont
  {Han}}, \bibinfo {author} {\bibfnamefont {F.}~\bibnamefont {Chen}}, \bibinfo
  {author} {\bibfnamefont {Z.}~\bibnamefont {Zhong}}, \bibinfo {author}
  {\bibfnamefont {K.}~\bibnamefont {Ramesh}}, \bibinfo {author} {\bibfnamefont
  {L.}~\bibnamefont {Chen}}, \ and\ \bibinfo {author} {\bibfnamefont
  {E.}~\bibnamefont {Widjaja}},\ }\href@noop {} {\bibfield  {journal} {\bibinfo
   {journal} {J. Phys. Chem. B}\ }\textbf {\bibinfo {volume} {110}},\ \bibinfo
  {pages} {24450} (\bibinfo {year} {2006})}\BibitemShut {NoStop}%
\bibitem [{\citenamefont {Jeong}\ \emph {et~al.}(2015)\citenamefont {Jeong},
  \citenamefont {Jin}, \citenamefont {Jerng}, \citenamefont {Seo},
  \citenamefont {Kim}, \citenamefont {Nahm}, \citenamefont {Kim},\ and\
  \citenamefont {Nam}}]{jeong15}%
  \BibitemOpen
  \bibfield  {author} {\bibinfo {author} {\bibfnamefont {D.}~\bibnamefont
  {Jeong}}, \bibinfo {author} {\bibfnamefont {K.}~\bibnamefont {Jin}}, \bibinfo
  {author} {\bibfnamefont {S.~E.}\ \bibnamefont {Jerng}}, \bibinfo {author}
  {\bibfnamefont {H.}~\bibnamefont {Seo}}, \bibinfo {author} {\bibfnamefont
  {D.}~\bibnamefont {Kim}}, \bibinfo {author} {\bibfnamefont {S.~H.}\
  \bibnamefont {Nahm}}, \bibinfo {author} {\bibfnamefont {S.~H.}\ \bibnamefont
  {Kim}}, \ and\ \bibinfo {author} {\bibfnamefont {K.~T.}\ \bibnamefont
  {Nam}},\ }\href@noop {} {\bibfield  {journal} {\bibinfo  {journal} {ACS
  Catalysis}\ }\textbf {\bibinfo {volume} {5}},\ \bibinfo {pages} {4624}
  (\bibinfo {year} {2015})}\BibitemShut {NoStop}%
\bibitem [{\citenamefont {Vidya}\ \emph {et~al.}(2006)\citenamefont {Vidya},
  \citenamefont {Ravindran}, \citenamefont {Fjellv{\aa}g},\ and\ \citenamefont
  {Kjekshus}}]{vidya2006}%
  \BibitemOpen
  \bibfield  {author} {\bibinfo {author} {\bibfnamefont {R.}~\bibnamefont
  {Vidya}}, \bibinfo {author} {\bibfnamefont {P.}~\bibnamefont {Ravindran}},
  \bibinfo {author} {\bibfnamefont {H.}~\bibnamefont {Fjellv{\aa}g}}, \ and\
  \bibinfo {author} {\bibfnamefont {A.}~\bibnamefont {Kjekshus}},\ }\href@noop
  {} {\bibfield  {journal} {\bibinfo  {journal} {Phys. Rev. B}\ }\textbf
  {\bibinfo {volume} {74}},\ \bibinfo {pages} {054422} (\bibinfo {year}
  {2006})}\BibitemShut {NoStop}%
\bibitem [{\citenamefont {Ansell}\ \emph {et~al.}(1982)\citenamefont {Ansell},
  \citenamefont {Modrick}, \citenamefont {Longo}, \citenamefont
  {Poeppeimeler},\ and\ \citenamefont {Horowitz}}]{ansell82}%
  \BibitemOpen
  \bibfield  {author} {\bibinfo {author} {\bibfnamefont {G.~B.}\ \bibnamefont
  {Ansell}}, \bibinfo {author} {\bibfnamefont {M.~A.}\ \bibnamefont {Modrick}},
  \bibinfo {author} {\bibfnamefont {J.}~\bibnamefont {Longo}}, \bibinfo
  {author} {\bibfnamefont {K.}~\bibnamefont {Poeppeimeler}}, \ and\ \bibinfo
  {author} {\bibfnamefont {H.}~\bibnamefont {Horowitz}},\ }\href@noop {}
  {\bibfield  {journal} {\bibinfo  {journal} {Acta Crystallogr. Sec. B}\
  }\textbf {\bibinfo {volume} {38}},\ \bibinfo {pages} {1795} (\bibinfo {year}
  {1982})}\BibitemShut {NoStop}%
\bibitem [{\citenamefont {Leceref}(1974)}]{lecerf74}%
  \BibitemOpen
  \bibfield  {author} {\bibinfo {author} {\bibfnamefont {A.}~\bibnamefont
  {Leceref}},\ }\href@noop {} {\bibfield  {journal} {\bibinfo  {journal} {C. R.
  Acad. Sci., Paris S\'{e}r. C}\ }\textbf {\bibinfo {volume} {279}},\ \bibinfo
  {pages} {879} (\bibinfo {year} {1974})}\BibitemShut {NoStop}%
\bibitem [{\citenamefont {Riou}\ and\ \citenamefont {Lecerf}(1977)}]{riou77}%
  \BibitemOpen
  \bibfield  {author} {\bibinfo {author} {\bibfnamefont {A.}~\bibnamefont
  {Riou}}\ and\ \bibinfo {author} {\bibfnamefont {A.}~\bibnamefont {Lecerf}},\
  }\href@noop {} {\bibfield  {journal} {\bibinfo  {journal} {Acta Crystallogr.
  Sec. B}\ }\textbf {\bibinfo {volume} {33}},\ \bibinfo {pages} {1896}
  (\bibinfo {year} {1977})}\BibitemShut {NoStop}%
\bibitem [{\citenamefont {Ravindran}\ \emph {et~al.}(2006)\citenamefont
  {Ravindran}, \citenamefont {Vidya}, \citenamefont {Kjekshus}, \citenamefont
  {Fjellv{\aa}g},\ and\ \citenamefont {Eriksson}}]{ravi2006}%
  \BibitemOpen
  \bibfield  {author} {\bibinfo {author} {\bibfnamefont {P.}~\bibnamefont
  {Ravindran}}, \bibinfo {author} {\bibfnamefont {R.}~\bibnamefont {Vidya}},
  \bibinfo {author} {\bibfnamefont {A.}~\bibnamefont {Kjekshus}}, \bibinfo
  {author} {\bibfnamefont {H.}~\bibnamefont {Fjellv{\aa}g}}, \ and\ \bibinfo
  {author} {\bibfnamefont {O.}~\bibnamefont {Eriksson}},\ }\href@noop {}
  {\bibfield  {journal} {\bibinfo  {journal} {Phys. Rev. B}\ }\textbf {\bibinfo
  {volume} {74}},\ \bibinfo {pages} {224412} (\bibinfo {year}
  {2006})}\BibitemShut {NoStop}%
\bibitem [{\citenamefont {Blaha}\ \emph {et~al.}(2001)\citenamefont {Blaha},
  \citenamefont {Schwarz}, \citenamefont {Madsen}, \citenamefont {Kvasnicka},\
  and\ \citenamefont {Luitz}}]{blaha01}%
  \BibitemOpen
  \bibfield  {author} {\bibinfo {author} {\bibfnamefont {P.}~\bibnamefont
  {Blaha}}, \bibinfo {author} {\bibfnamefont {K.}~\bibnamefont {Schwarz}},
  \bibinfo {author} {\bibfnamefont {G.}~\bibnamefont {Madsen}}, \bibinfo
  {author} {\bibfnamefont {D.}~\bibnamefont {Kvasnicka}}, \ and\ \bibinfo
  {author} {\bibfnamefont {J.}~\bibnamefont {Luitz}},\ }\href@noop {}
  {\bibfield  {journal} {\bibinfo  {journal} {An augmented plane wave+ local
  orbitals program for calculating crystal properties}\ } (\bibinfo {year}
  {2001})}\BibitemShut {NoStop}%
\bibitem [{\citenamefont {Perdew}\ \emph {et~al.}(1996)\citenamefont {Perdew},
  \citenamefont {Burke},\ and\ \citenamefont {Ernzerhof}}]{perdew96}%
  \BibitemOpen
  \bibfield  {author} {\bibinfo {author} {\bibfnamefont {J.}~\bibnamefont
  {Perdew}}, \bibinfo {author} {\bibfnamefont {K.}~\bibnamefont {Burke}}, \
  and\ \bibinfo {author} {\bibfnamefont {M.}~\bibnamefont {Ernzerhof}},\
  }\href@noop {} {\bibfield  {journal} {\bibinfo  {journal} {Phys. Rev. Lett.}\
  }\textbf {\bibinfo {volume} {78}},\ \bibinfo {pages} {1396} (\bibinfo {year}
  {1996})}\BibitemShut {NoStop}%
\bibitem [{\citenamefont {Kresse}\ and\ \citenamefont
  {Furthm{\"u}ller}(1996)}]{kresse96}%
  \BibitemOpen
  \bibfield  {author} {\bibinfo {author} {\bibfnamefont {G.}~\bibnamefont
  {Kresse}}\ and\ \bibinfo {author} {\bibfnamefont {J.}~\bibnamefont
  {Furthm{\"u}ller}},\ }\href@noop {} {\bibfield  {journal} {\bibinfo
  {journal} {Computational Materials Science}\ }\textbf {\bibinfo {volume}
  {6}},\ \bibinfo {pages} {15} (\bibinfo {year} {1996})}\BibitemShut {NoStop}%
\bibitem [{\citenamefont {Gao}\ \emph {et~al.}(2009{\natexlab{b}})\citenamefont
  {Gao}, \citenamefont {Krumeich}, \citenamefont {Nesper}, \citenamefont
  {Fjellv{\aa}g},\ and\ \citenamefont {Norby}}]{gao2009}%
  \BibitemOpen
  \bibfield  {author} {\bibinfo {author} {\bibfnamefont {T.}~\bibnamefont
  {Gao}}, \bibinfo {author} {\bibfnamefont {F.}~\bibnamefont {Krumeich}},
  \bibinfo {author} {\bibfnamefont {R.}~\bibnamefont {Nesper}}, \bibinfo
  {author} {\bibfnamefont {H.}~\bibnamefont {Fjellv{\aa}g}}, \ and\ \bibinfo
  {author} {\bibfnamefont {P.}~\bibnamefont {Norby}},\ }\href@noop {}
  {\bibfield  {journal} {\bibinfo  {journal} {Inorg. Chem.}\ }\textbf {\bibinfo
  {volume} {48}},\ \bibinfo {pages} {6242} (\bibinfo {year}
  {2009}{\natexlab{b}})}\BibitemShut {NoStop}%
\bibitem [{\citenamefont {Hauback}\ \emph {et~al.}(2000)\citenamefont
  {Hauback}, \citenamefont {Fjellv{\aa}g}, \citenamefont {Steinsvoll},
  \citenamefont {Johansson}, \citenamefont {Buset},\ and\ \citenamefont
  {J{\o}rgensen}}]{hauback00}%
  \BibitemOpen
  \bibfield  {author} {\bibinfo {author} {\bibfnamefont {B.~C.}\ \bibnamefont
  {Hauback}}, \bibinfo {author} {\bibfnamefont {H.}~\bibnamefont
  {Fjellv{\aa}g}}, \bibinfo {author} {\bibfnamefont {O.}~\bibnamefont
  {Steinsvoll}}, \bibinfo {author} {\bibfnamefont {K.}~\bibnamefont
  {Johansson}}, \bibinfo {author} {\bibfnamefont {O.~T.}\ \bibnamefont
  {Buset}}, \ and\ \bibinfo {author} {\bibfnamefont {J.}~\bibnamefont
  {J{\o}rgensen}},\ }\href@noop {} {\bibfield  {journal} {\bibinfo  {journal}
  {J. Neut. Res.}\ }\textbf {\bibinfo {volume} {8}},\ \bibinfo {pages} {215}
  (\bibinfo {year} {2000})}\BibitemShut {NoStop}%
\bibitem [{\citenamefont {Rietveld}(1969)}]{rietveld69}%
  \BibitemOpen
  \bibfield  {author} {\bibinfo {author} {\bibfnamefont {H.}~\bibnamefont
  {Rietveld}},\ }\href@noop {} {\bibfield  {journal} {\bibinfo  {journal} {J.
  Appl. Crystallogr.}\ }\textbf {\bibinfo {volume} {2}},\ \bibinfo {pages} {65}
  (\bibinfo {year} {1969})}\BibitemShut {NoStop}%
\bibitem [{\citenamefont {Rodr{\'\i}guez-Carvajal}(1993)}]{rodriguez93}%
  \BibitemOpen
  \bibfield  {author} {\bibinfo {author} {\bibfnamefont {J.}~\bibnamefont
  {Rodr{\'\i}guez-Carvajal}},\ }\href@noop {} {\bibfield  {journal} {\bibinfo
  {journal} {Physica B}\ }\textbf {\bibinfo {volume} {192}},\ \bibinfo {pages}
  {55} (\bibinfo {year} {1993})}\BibitemShut {NoStop}%
\bibitem [{\citenamefont {Oswald}\ \emph {et~al.}(1965)\citenamefont {Oswald},
  \citenamefont {Feitknecht},\ and\ \citenamefont {Wampetich}}]{oswald65}%
  \BibitemOpen
  \bibfield  {author} {\bibinfo {author} {\bibfnamefont {H.}~\bibnamefont
  {Oswald}}, \bibinfo {author} {\bibfnamefont {W.}~\bibnamefont {Feitknecht}},
  \ and\ \bibinfo {author} {\bibfnamefont {M.}~\bibnamefont {Wampetich}},\
  }\href@noop {} {\bibfield  {journal} {\bibinfo  {journal} {Nature}\ }\textbf
  {\bibinfo {volume} {207}},\ \bibinfo {pages} {72} (\bibinfo {year}
  {1965})}\BibitemShut {NoStop}%
\bibitem [{\citenamefont {Haines}\ \emph {et~al.}(1995)\citenamefont {Haines},
  \citenamefont {Leger},\ and\ \citenamefont {Hoyau}}]{haines95}%
  \BibitemOpen
  \bibfield  {author} {\bibinfo {author} {\bibfnamefont {J.}~\bibnamefont
  {Haines}}, \bibinfo {author} {\bibfnamefont {J.~M.}\ \bibnamefont {Leger}}, \
  and\ \bibinfo {author} {\bibfnamefont {S.}~\bibnamefont {Hoyau}},\
  }\href@noop {} {\bibfield  {journal} {\bibinfo  {journal} {J. Phys. Chem.
  Sol.}\ }\textbf {\bibinfo {volume} {56}},\ \bibinfo {pages} {965} (\bibinfo
  {year} {1995})}\BibitemShut {NoStop}%
\bibitem [{\citenamefont {Jeanloz}\ and\ \citenamefont
  {Rudy}(1987)}]{jeanloz87}%
  \BibitemOpen
  \bibfield  {author} {\bibinfo {author} {\bibfnamefont {R.}~\bibnamefont
  {Jeanloz}}\ and\ \bibinfo {author} {\bibfnamefont {A.}~\bibnamefont {Rudy}},\
  }\href@noop {} {\bibfield  {journal} {\bibinfo  {journal} {Journal of
  Geophysical Research: Solid Earth}\ }\textbf {\bibinfo {volume} {92}},\
  \bibinfo {pages} {11433} (\bibinfo {year} {1987})}\BibitemShut {NoStop}%
\bibitem [{\citenamefont {Cohen}\ \emph {et~al.}(1997)\citenamefont {Cohen},
  \citenamefont {Mazin},\ and\ \citenamefont {Isaak}}]{cohen97}%
  \BibitemOpen
  \bibfield  {author} {\bibinfo {author} {\bibfnamefont {R.~E.}\ \bibnamefont
  {Cohen}}, \bibinfo {author} {\bibfnamefont {I.}~\bibnamefont {Mazin}}, \ and\
  \bibinfo {author} {\bibfnamefont {D.~G.}\ \bibnamefont {Isaak}},\ }\href@noop
  {} {\bibfield  {journal} {\bibinfo  {journal} {Science}\ }\textbf {\bibinfo
  {volume} {275}},\ \bibinfo {pages} {654} (\bibinfo {year}
  {1997})}\BibitemShut {NoStop}%
\bibitem [{\citenamefont {Ravindran}\ \emph {et~al.}(2001)\citenamefont
  {Ravindran}, \citenamefont {Kjekshus}, \citenamefont {Fjellv{\aa}g},
  \citenamefont {James}, \citenamefont {Nordstr{\"o}m}, \citenamefont
  {Johansson},\ and\ \citenamefont {Eriksson}}]{ravindran01}%
  \BibitemOpen
  \bibfield  {author} {\bibinfo {author} {\bibfnamefont {P.}~\bibnamefont
  {Ravindran}}, \bibinfo {author} {\bibfnamefont {A.}~\bibnamefont {Kjekshus}},
  \bibinfo {author} {\bibfnamefont {H.}~\bibnamefont {Fjellv{\aa}g}}, \bibinfo
  {author} {\bibfnamefont {P.}~\bibnamefont {James}}, \bibinfo {author}
  {\bibfnamefont {L.}~\bibnamefont {Nordstr{\"o}m}}, \bibinfo {author}
  {\bibfnamefont {B.}~\bibnamefont {Johansson}}, \ and\ \bibinfo {author}
  {\bibfnamefont {O.}~\bibnamefont {Eriksson}},\ }\href@noop {} {\bibfield
  {journal} {\bibinfo  {journal} {Phys. Rev. B}\ }\textbf {\bibinfo {volume}
  {63}},\ \bibinfo {pages} {144409} (\bibinfo {year} {2001})}\BibitemShut
  {NoStop}%
\bibitem [{\citenamefont {Brooks}(1985)}]{brooks85}%
  \BibitemOpen
  \bibfield  {author} {\bibinfo {author} {\bibfnamefont {M.}~\bibnamefont
  {Brooks}},\ }\href@noop {} {\bibfield  {journal} {\bibinfo  {journal}
  {Physica B}\ }\textbf {\bibinfo {volume} {130}},\ \bibinfo {pages} {6}
  (\bibinfo {year} {1985})}\BibitemShut {NoStop}%
\bibitem [{\citenamefont {Eriksson}\ \emph {et~al.}(1989)\citenamefont
  {Eriksson}, \citenamefont {Johansson},\ and\ \citenamefont
  {Brooks}}]{eriksson89}%
  \BibitemOpen
  \bibfield  {author} {\bibinfo {author} {\bibfnamefont {O.}~\bibnamefont
  {Eriksson}}, \bibinfo {author} {\bibfnamefont {B.}~\bibnamefont {Johansson}},
  \ and\ \bibinfo {author} {\bibfnamefont {M.}~\bibnamefont {Brooks}},\
  }\href@noop {} {\bibfield  {journal} {\bibinfo  {journal} {J. Phys. Condens.
  Matter}\ }\textbf {\bibinfo {volume} {1}},\ \bibinfo {pages} {4005} (\bibinfo
  {year} {1989})}\BibitemShut {NoStop}%
\bibitem [{\citenamefont {Goodenough}()}]{goodenough59}%
  \BibitemOpen
  \bibfield  {author} {\bibinfo {author} {\bibfnamefont {J.}~\bibnamefont
  {Goodenough}},\ }\href@noop {} {\bibinfo  {journal} {Magnetism and the
  Chemical Bond (Wiley, New York, 1963)}\ }\BibitemShut {NoStop}%
\bibitem [{\citenamefont {Motida}\ and\ \citenamefont
  {Miyahara}(1970)}]{motida90}%
  \BibitemOpen
\bibfield  {journal} {  }\bibfield  {author} {\bibinfo {author} {\bibfnamefont
  {K.}~\bibnamefont {Motida}}\ and\ \bibinfo {author} {\bibfnamefont
  {S.}~\bibnamefont {Miyahara}},\ }\href@noop {} {\bibfield  {journal}
  {\bibinfo  {journal} {J. Phys. Soc. Jpn}\ }\textbf {\bibinfo {volume} {28}},\
  \bibinfo {pages} {1188} (\bibinfo {year} {1970})}\BibitemShut {NoStop}%
\bibitem [{\citenamefont {Kanamori}(1959)}]{kanamori59}%
  \BibitemOpen
  \bibfield  {author} {\bibinfo {author} {\bibfnamefont {J.}~\bibnamefont
  {Kanamori}},\ }\href@noop {} {\bibfield  {journal} {\bibinfo  {journal} {J.
  Phys. Chem. Solids}\ }\textbf {\bibinfo {volume} {10}},\ \bibinfo {pages}
  {87} (\bibinfo {year} {1959})}\BibitemShut {NoStop}%
\bibitem [{\citenamefont {Deringer}\ \emph {et~al.}(2011)\citenamefont
  {Deringer}, \citenamefont {Tchougr{\'e}eff},\ and\ \citenamefont
  {Dronskowski}}]{deringer11}%
  \BibitemOpen
  \bibfield  {author} {\bibinfo {author} {\bibfnamefont {V.~L.}\ \bibnamefont
  {Deringer}}, \bibinfo {author} {\bibfnamefont {A.~L.}\ \bibnamefont
  {Tchougr{\'e}eff}}, \ and\ \bibinfo {author} {\bibfnamefont {R.}~\bibnamefont
  {Dronskowski}},\ }\href@noop {} {\bibfield  {journal} {\bibinfo  {journal}
  {J. Phys. Chem. A}\ }\textbf {\bibinfo {volume} {115}},\ \bibinfo {pages}
  {5461} (\bibinfo {year} {2011})}\BibitemShut {NoStop}%
\bibitem [{\citenamefont {Maintz}\ \emph {et~al.}(2013)\citenamefont {Maintz},
  \citenamefont {Deringer}, \citenamefont {Tchougr{\'e}eff},\ and\
  \citenamefont {Dronskowski}}]{maintz13}%
  \BibitemOpen
  \bibfield  {author} {\bibinfo {author} {\bibfnamefont {S.}~\bibnamefont
  {Maintz}}, \bibinfo {author} {\bibfnamefont {V.~L.}\ \bibnamefont
  {Deringer}}, \bibinfo {author} {\bibfnamefont {A.~L.}\ \bibnamefont
  {Tchougr{\'e}eff}}, \ and\ \bibinfo {author} {\bibfnamefont {R.}~\bibnamefont
  {Dronskowski}},\ }\href@noop {} {\bibfield  {journal} {\bibinfo  {journal}
  {J. Comput.Chem.}\ }\textbf {\bibinfo {volume} {34}},\ \bibinfo {pages}
  {2557} (\bibinfo {year} {2013})}\BibitemShut {NoStop}%
\bibitem [{\citenamefont {Dronskowski}\ and\ \citenamefont
  {Bloechl}(1993)}]{dronskowski93}%
  \BibitemOpen
  \bibfield  {author} {\bibinfo {author} {\bibfnamefont {R.}~\bibnamefont
  {Dronskowski}}\ and\ \bibinfo {author} {\bibfnamefont {P.~E.}\ \bibnamefont
  {Bloechl}},\ }\href@noop {} {\bibfield  {journal} {\bibinfo  {journal} {J.
  Phys. Chem.}\ }\textbf {\bibinfo {volume} {97}},\ \bibinfo {pages} {8617}
  (\bibinfo {year} {1993})}\BibitemShut {NoStop}%
\bibitem [{\citenamefont {Seo}\ \emph {et~al.}(2004)\citenamefont {Seo},
  \citenamefont {Jo}, \citenamefont {Lee}, \citenamefont {Kim}, \citenamefont
  {Oh},\ and\ \citenamefont {Park}}]{seo04}%
  \BibitemOpen
  \bibfield  {author} {\bibinfo {author} {\bibfnamefont {W.~S.}\ \bibnamefont
  {Seo}}, \bibinfo {author} {\bibfnamefont {H.~H.}\ \bibnamefont {Jo}},
  \bibinfo {author} {\bibfnamefont {K.}~\bibnamefont {Lee}}, \bibinfo {author}
  {\bibfnamefont {B.}~\bibnamefont {Kim}}, \bibinfo {author} {\bibfnamefont
  {S.~J.}\ \bibnamefont {Oh}}, \ and\ \bibinfo {author} {\bibfnamefont {J.~T.}\
  \bibnamefont {Park}},\ }\href@noop {} {\bibfield  {journal} {\bibinfo
  {journal} {Angew. Chem. Int. Ed.}\ }\textbf {\bibinfo {volume} {43}},\
  \bibinfo {pages} {1115} (\bibinfo {year} {2004})}\BibitemShut {NoStop}%
\bibitem [{\citenamefont {Bertaut}(1968)}]{bertaut68}%
  \BibitemOpen
  \bibfield  {author} {\bibinfo {author} {\bibfnamefont {E.}~\bibnamefont
  {Bertaut}},\ }\href@noop {} {\bibfield  {journal} {\bibinfo  {journal} {Acta
  Crystallogr. Sec. A}\ }\textbf {\bibinfo {volume} {24}},\ \bibinfo {pages}
  {217} (\bibinfo {year} {1968})}\BibitemShut {NoStop}%
\bibitem [{\citenamefont {Kovalev}(1985)}]{kovalev85}%
  \BibitemOpen
  \bibfield  {author} {\bibinfo {author} {\bibfnamefont {O.~V.}\ \bibnamefont
  {Kovalev}},\ }\href@noop {} {\emph {\bibinfo {title} {Irreducible
  representations of the space groups}}}\ (\bibinfo  {publisher} {Gordon and
  Breach, New York},\ \bibinfo {year} {1985})\BibitemShut {NoStop}%
\end{thebibliography}%

\end{document}